\PassOptionsToPackage{table,dvipsnames,xcdraw}{xcolor}
\documentclass[sigconf,screen]{acmart}
\setcopyright{acmlicensed}
\acmPrice{15.00}
\acmDOI{10.1145/3540250.3549147}
\acmYear{2022}
\copyrightyear{2022}
\acmSubmissionID{fse22main-p777-p}
\acmISBN{978-1-4503-9413-0/22/11}
\acmConference[ESEC/FSE '22]{Proceedings of the 30th ACM Joint European Software Engineering Conference and Symposium on the Foundations of Software Engineering}{November 14--18, 2022}{Singapore, Singapore}
\acmBooktitle{Proceedings of the 30th ACM Joint European Software Engineering Conference and Symposium on the Foundations of Software Engineering (ESEC/FSE '22), November 14--18, 2022, Singapore, Singapore}
\usepackage{multirow}

\usepackage[font=small,labelfont=bf,tableposition=top]{caption}
\usepackage{mathtools}

\usepackage{colortbl}
  
\usepackage{mdwmath}

\usepackage{algorithm}
\usepackage{array}
\usepackage{hhline}
\usepackage{environ}
\usepackage[noend]{algpseudocode}
\usepackage{booktabs}
\usepackage{wrapfig}

\usepackage{color}
\usepackage[dvipsnames]{xcolor}
\usepackage{enumitem}
\usepackage{url}

\usepackage{stfloats}

\usepackage{tabularx}
\usepackage{rotating}
\usepackage{times}
\usepackage{comment}
\usepackage{amsthm}
\usepackage{dsfont}
\usepackage{pifont}
\usepackage{flushend}
\usepackage{makecell}
\usepackage{xspace} %for leaving a space after a string macro, see http below
\usepackage{setspace}
\usepackage{bm}
\usepackage[export]{adjustbox}
\usepackage{stfloats} % put table at bottom
\usepackage{microtype} % beautify
\usepackage{diagbox}
\usepackage{mathpartir}
\usepackage[framemethod=TikZ]{mdframed}

\usepackage{listings}
\definecolor{codegreen}{rgb}{0,0.6,0}
\definecolor{codegray}{rgb}{0.5,0.5,0.5}
\definecolor{codepurple}{rgb}{0.58,0,0.82}
\definecolor{backcolour}{rgb}{0.95,0.95,0.92}
\lstdefinelanguage
   [x64]{Assembler}     % add a "x64" dialect of Assembler
   [x86masm]{Assembler} % based on the "x86masm" dialect
   {morekeywords={CDQE,CQO,CMPSQ,CMPXCHG16B,JRCXZ,LODSQ,MOVSXD, %
                  POPFQ,PUSHFQ,SCASQ,STOSQ,IRETQ,RDTSCP,SWAPGS, %
                  rax,rdx,rcx,rbx,rsi,rdi,rsp,rbp,rip %
                  r8,r8d,r8w,r8b,r9,r9d,r9w,r9b, %
                  r10,r10d,r10w,r10b,r11,r11d,r11w,r11b, %
                  r12,r12d,r12w,r12b,r13,r13d,r13w,r13b, %
                  r14,r14d,r14w,r14b,r15,r15d,r15w,r15b},
    keywordstyle=\ttfamily\bfseries\color{code}} % etc.
\lstdefinestyle{mystyle}{  
    commentstyle=\color{codegreen},
    numberstyle=\footnotesize\color{black},
    basicstyle=\ttfamily\bfseries,
    breakatwhitespace=false,         
    breaklines=true,                 
    captionpos=b,                    
    keepspaces=true, 
    keywordstyle=\color{code}
    numbersep=5pt,                  
    showspaces=false,                
    showstringspaces=false,
    showtabs=false,
    stepnumber=1,
    tabsize=2,
}
\usepackage[caption=false,font=footnotesize]{subfig}

\usepackage{tikz}
\newmdenv[
    skipabove=0.5em,
    skipbelow=0.5em,
    linewidth=0.8pt,
    innerleftmargin=4pt,
    innerrightmargin=4pt,
    innertopmargin=2pt,
    innerbottommargin=2pt,
    linecolor=gray95,
    roundcorner=2pt, 
]{questionbox}

\newenvironment{question}
{\begin{questionbox}}
{\end{questionbox}}

\microtypecontext{spacing=nonfrench}

\newcommand{\ie}{\emph{i.e., }}
\newcommand{\eg}{\emph{e.g., }}

\newcommand{\code}[1]{\texttt{\color{code}\textbf{#1}}}
\newcommand{\comments}[1]{\texttt{\color{codegreen}#1}}
\newcommand{\mask}[1]{\colorbox{lightgray}{\texttt{\textbf{#1}}}}

\newtheoremstyle{mydef}
  {.1cm}   % ABOVESPACE
  {0cm}   % BELOWSPACE
  {\normalfont}  % BODYFONT
  {0pt}       % INDENT (empty value is the same as 0pt)
  {\bfseries} % HEADFONT
  {.}         % HEADPUNCT
  {5pt plus 1pt minus 1pt} % HEADSPACE
  {}          % CUSTOM-HEAD-SPEC
  
\theoremstyle{mydef}
\newtheorem{definition}{Definition}[section]
\newtheorem{example}{Example}[section]

\newcommand{\para}[1]{\vspace{.1cm}\noindent\textbf{\emph{#1}}}

\definecolor{Gray}{gray}{0.85}
\definecolor{myblue}{rgb}{0,.6,1}
\definecolor{darkgray}{rgb}{.4,.4,.4}
\definecolor{lightblue}{HTML}{DAE8FC}
\definecolor{lightred}{HTML}{F8CECC}
\definecolor{lightgray}{HTML}{B3B3B3}
\definecolor{shadecolor}{HTML}{DCDCDC}
\definecolor{lightred}{HTML}{FF9999}
\definecolor{lightorange}{HTML}{FFCC99}
\definecolor{gray95}{gray}{0.05}
\definecolor{code}{HTML}{660066}
\definecolor{ao(english)}{rgb}{0.0, 0.5, 0.0}

\newcolumntype{?}[1]{!{\vrule width #1}}

\def\sys{\textsc{NeuDep}\xspace}

\newcommand{\etal}{\textit{et al.~}}

\DeclareMathOperator*{\argmin}{arg\,min}

\usepackage{booktabs}   %% For formal tables:
\begin{document}\sloppy

%% Title information
\title[\sys]{\sys: Neural Binary Memory Dependence Analysis}

\author{Kexin Pei}
\email{kpei@cs.columbia.edu}
\affiliation{%
  \institution{Columbia University}
  \city{New York}
  \country{USA}}

\author{Dongdong She}
\authornote{These authors contributed equally to the paper}
\email{dongdong@cs.columbia.edu}
\affiliation{%
  \institution{Columbia University}
  \city{New York}
  \country{USA}
}

\author{Michael Wang}
\authornotemark[1]
\email{mi27950@mit.edu}
\affiliation{%
  \institution{MIT}
  \city{Cambridge}
  \country{USA}}

\author{Scott Geng}
\authornotemark[1]
\email{scott.geng@columbia.edu}
\affiliation{%
  \institution{Columbia University}
  \city{New York}
  \country{USA}}

\author{Zhou Xuan}
\email{xuan1@purdue.edu}
\affiliation{%
  \institution{Purdue University}
  \city{West Lafayette}
  \country{USA}}

\author{Yaniv David}
\email{yaniv.david@columbia.edu}
\affiliation{%
  \institution{Columbia University}
  \city{New York}
  \country{USA}}

\author{Junfeng Yang}
\email{junfeng@cs.columbia.edu}
\affiliation{%
  \institution{Columbia University}
  \city{New York}
  \country{USA}}

\author{Suman Jana}
\email{suman@cs.columbia.edu}
\affiliation{%
  \institution{Columbia University}
  \city{New York}
  \country{USA}}

\author{Baishakhi Ray}
\email{rayb@cs.columbia.edu}
\affiliation{%
  \institution{Columbia University}
  \city{New York}
  \country{USA}}

\renewcommand{\shortauthors}{Pei, She, Wang, Geng, Xuan, David, Yang, Jana, Ray}

\begin{abstract}

Determining whether multiple instructions can access the same memory location is a critical task in binary analysis. It is challenging as statically computing precise alias information is undecidable in theory. The problem aggravates at the binary level due to the presence of compiler optimizations and the absence of symbols and types. Existing approaches either produce significant spurious dependencies due to conservative analysis or scale poorly to complex binaries.

We present a new machine-learning-based approach to predict memory dependencies by exploiting the model's learned knowledge about how binary programs execute. Our approach features (i) a self-supervised procedure that pretrains a neural net to reason over binary code and its dynamic value flows through memory addresses, followed by (ii) supervised finetuning to infer the memory dependencies statically. To facilitate efficient learning, we develop dedicated neural architectures to encode the heterogeneous inputs (\ie code, data values, and memory addresses from traces) with specific modules and fuse them with a composition learning strategy.

We implement our approach in \sys and evaluate it on 41 popular software projects compiled by 2 compilers, 4 optimizations, and 4 obfuscation passes. We demonstrate that \sys is more precise (1.5$\times$) and faster (3.5$\times$) than the current state-of-the-art. Extensive probing studies on security-critical reverse engineering tasks suggest that \sys understands memory access patterns, learns function signatures, and is able to match indirect calls. All these tasks either assist or benefit from inferring memory dependencies. Notably, \sys also outperforms the current state-of-the-art on these tasks.

\end{abstract}




% make the title area
\maketitle

\section{Introduction}
\label{sec:intro}

Binary memory dependence analysis, which determines whether two machine instructions in an executable can access the same memory location, is critical for many security-sensitive tasks, including detecting vulnerabilities~\cite{hefreewill, wang2019oo7, cova2006static}, analyzing malware~\cite{hernandez2017firmusb, yin2007panorama}, hardening binaries~\cite{jain2022bird, erlingsson2006xfi, arras2022sabre, williams2020egalito}, and forensics~\cite{xu2017postmortem, cui2018rept, mu2019renn, guo2019vsa}. 
The key challenge behind memory dependence analysis is that machine instructions often leverage indirect addressing or indirect control-flow transfer (\ie involving dynamically computed targets) to access the memory. 
Furthermore, most commercial software is stripped of source-level information such as variables, arguments, types, data structures, etc. 
Without this information, the problem of memory dependence analysis becomes even harder, forcing the analysis to reason about values flowing through generic registers and memory addresses.
Consider the following code snippet where we show two instructions (within the same function) at different program locations.
The function is executed twice, resulting in two different traces.
\begin{flushleft}
\footnotesize
\setlength{\tabcolsep}{4pt}
\renewcommand{\arraystretch}{1.1}

\begin{tabular}{rlll}
\toprule[1pt]
Address & Instruction & Trace 1 & Trace 2 \\ \midrule[.9pt]
\multicolumn{2}{l}{......} & \\
\texttt{0x06:} & \code{mov [rax],rbx}$^*$ & \comments{\textbf{rax=0x3;rbx=\underline{0x1}}} & \comments{\textbf{rax=0x5;rbx=0x1}} \\
\multicolumn{2}{l}{......} & \\
\texttt{0x1f:} &  \code{mov rdi,[0x3]} & \comments{\textbf{rdi=\underline{0x1}}} & \comments{\textbf{rdi=0x0}} \\
\multicolumn{2}{l}{......} & \\
\bottomrule[1pt]
\multicolumn{4}{l}{\multirow{2}{*}{\begin{tabular}[c]{@{}l@{}}\scriptsize $^*$In Intel x86 syntax~\cite{studio2006x86}, \code{mov [rax],rbx} means writing register \code{rbx} to the memory\\ pointed by register \code{rax}; [] means dereference a memory address.\end{tabular}}} \\
\multicolumn{4}{l}{}\\

\end{tabular}
\end{flushleft}
The two instructions are memory-dependent (read-after-write) when \code{rax}=\texttt{0x3} (Trace 1).
When analyzing the code statically, it requires precise value flow analysis to determine what values can flow to \code{rax} from different program contexts.

Over the last two decades, researchers have made numerous attempts to improve the accuracy and performance of binary memory dependence analysis~\cite{debray1998alias, cifuentes1997intraprocedural, guo2005practical, brumley2006alias, balakrishnan2004vsa, balakrishnan2010wysinwyx, reps2008improved}.
The most common approach often involves statically computing and propagating an over-approximated set of values that each register and memory address can contain at each program point using abstract interpretation.
For example, a seminal paper by Balakrishnan and Reps on value set analysis (VSA)~\cite{balakrishnan2004vsa} adopts strided intervals as the abstract domain and propagates the interval bounds for the operands (\eg registers and memory locations) along each instruction. 
VSA detects two instructions to be dependent if their intervals intersect.
Unfortunately, these static approaches have been shown to be highly imprecise in practice~\cite{zhang2019bda}.
\emph{Composing} abstract domains along multiple instructions and \emph{merging} them across a large number of paths quickly accumulate prohibitive amounts of over-approximation error.
As a result, the computed set of accessed memory addresses by such approaches often ends up covering almost the entire memory space, leading to a large number of false positives (\ie instructions with no dependencies are incorrectly detected as dependent).

With the advent of data-driven approaches to program analyses~\cite{pradel2021neural, wang2020blended, allamanis2018survey}, state-of-the-art memory dependence analysis is increasingly using statistical or machine learning (ML) based methods to improve the analysis precision~\cite{zhang2019bda, guo2019vsa, wang2018spindle, mu2019renn}, but they still suffer from serious limitations.
For example, DeepVSA~\cite{guo2019vsa} trains a neural network on static code to classify the memory locations accessed by each instruction into a more coarse-grained abstract domain such as stack, heap, and global, and use the predicted memory region to instantiate the value set in VSA.
However, such coarse-grained prediction results in high false positives as any two instructions accessing the same region (\eg stack) will always be detected as dependent even when the instructions access two completely different addresses. 
To avoid the precision losses by the static approaches, BDA~\cite{zhang2019bda} uses a dynamic approach that leverages probabilistic analysis to sample program paths and performs per-path abstract interpretation.
However, as real-world programs often have many paths, the cost of performing per-path abstract interpretation for even a smaller subset of paths adds prohibitive runtime overhead, \eg taking more than 12 hours to finish analyzing a single program.
It is perhaps not surprising---while a dynamic approach can be more accurate than static approaches, it can incur extremely high runtime overhead, especially while trying to achieve good code coverage. 

\begin{figure}[!t]
\centering
\includegraphics[width=\linewidth]{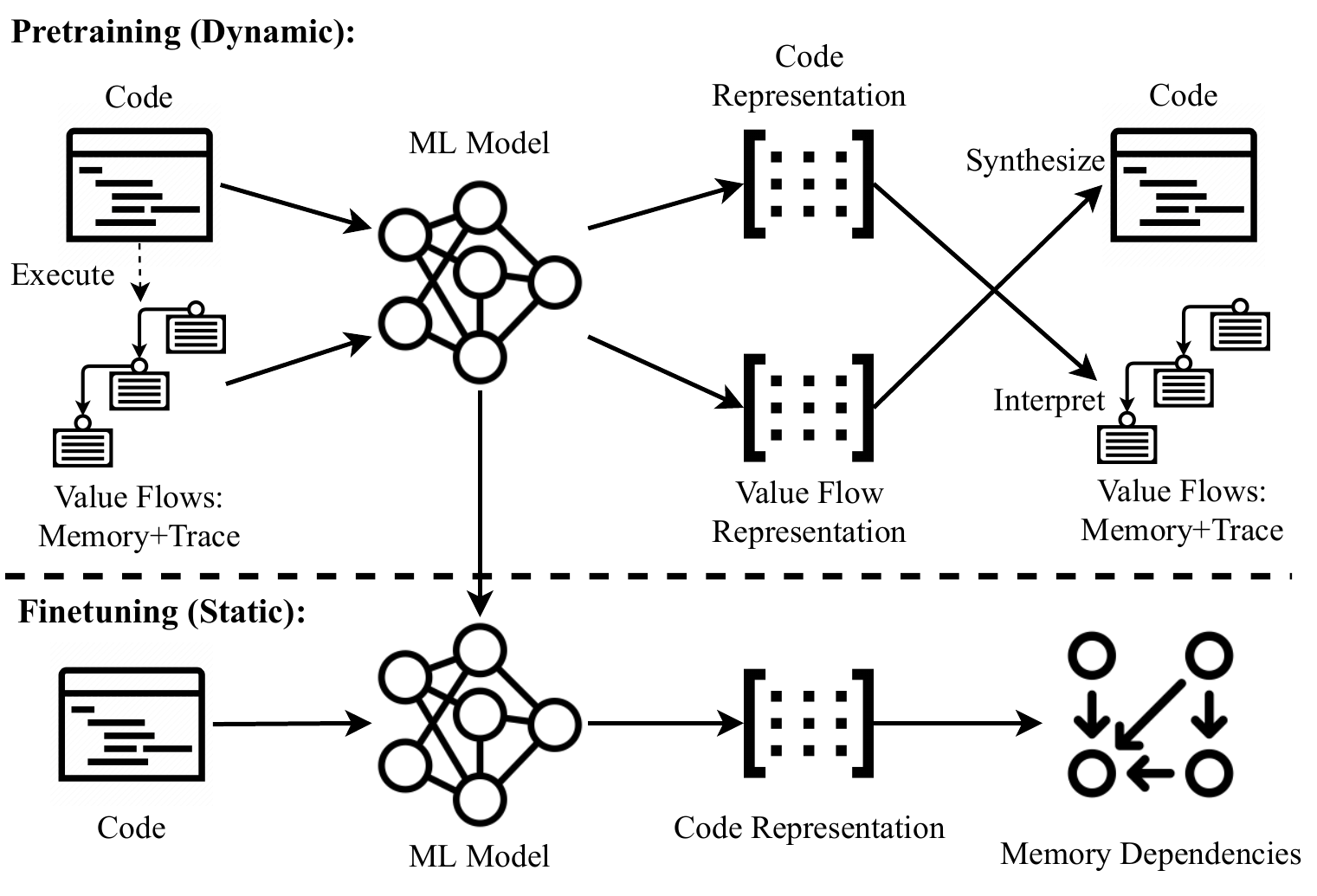}

\caption{The workflow of our approach. We first pretrain the model to predict code based on its traces and predict traces based on its code. We then finetune the model to statically infer memory dependencies.
% ~\bray{explicitly add memory address in the memory to show multi-modality}
}

\label{fig:workflow}
\end{figure}

To achieve higher accuracy in a reasonably faster time, we propose an ML-based hybrid approach. 
Our key strategy is to learn to reason about approximate memory dependencies from the execution behavior of generic binary code during training. 
We then apply the learned knowledge to static code during inference without any extra runtime overhead (see Figure~\ref{fig:workflow}). 
Such a hybrid approach, \ie learning from both code and traces, has been shown promise in several Software Engineering applications, including clone detection, type inference, and program fixing and synthesis~\cite{pei2021stateformer, patra2021nalin, nye2021show, wang2017dynamic, wang2020blended}. 
However, none of these works can reason fine-grained value flows through different memory addresses as they do not explicitly model memory. 
To bridge this gap, we aim to model the memory addresses in the ML-based hybrid framework and try to make fine-grained predictions differentiating the memory contents of different data pointers.
Modeling memory address is, however, challenging as it requires the model to (i) distinguish between different memory addresses, (ii) learn to reason about indirect address references and memory contents, and (iii) learn the compositional effects of multiple instructions that involve memory operations.

To this end, we propose a new learning framework comprising pretraining and finetuning steps inspired by masked language model (MLM)~\cite{devlin2018bert}, as shown in Figure~\ref{fig:workflow}. 
Unlike traditional MLM, where the input is restricted to a single input modality (\eg text), our model learns from multi-modal information: instructions (static code), traces (dynamic values), and memory addresses (code spatial layout). 
We deploy a novel fusion module to simultaneously capture the interactions of these modalities for predicting memory dependencies. 
During pretraining, we mask random tokens from these modalities.  
While predicting masked opcode teaches the model to {\em synthesize} the instruction, predicting masked values in traces and memory addresses forces the model to learn to {\em interpret} instructions and their effect on registers and memory contents. 
For instance, if we mask the value of \code{rax} in \code{mov [rax],rbx} in the above example and train the model to predict it, the model is forced to interpret the previous instructions in the context and reason about how they compute their trace values that flow into \code{rax}.
We hypothesize that such pretraining helps the model gain a general understanding of the value flow behavior involving memory operations.

After pretraining, the model is finetuned to statically (without the trace values) reason about the value flows (based on its learned knowledge from pretraining) across memory along multiple paths and predict the memory-dependent instruction pairs. 
Both pretraining and finetuning steps are automated and data-driven without manually defining any propagation rules for value flows. 
As a result, we show that our model is faster and more precise than the state-of-the-art systems (\S\ref{sec:eval}).

We implement our approach in \sys by carefully designing a new neural architecture specialized for fine-grained modeling of pointers to distinguish between unique memory addresses (Challenge i). 
We develop a novel fusion module to facilitate efficient training on the multi-modal bi-directional masking task, which helps the model to understand memory address content and thus, indirect memory references (Challenge ii). 
Finally, to teach the \emph{compositional effects} of instructions on memory values (Challenge iii), we leverage the principle of curriculum learning~\cite{bengio2009curriculum}, \ie expose short training examples in the initial learning phase, and gradually increase the sample difficulties as the training progresses. 

We evaluate \sys on a wide range of popular software projects compiled with diverse optimizations and obfuscation passes. 
We demonstrate that \sys is significantly more precise than state-of-the-art binary dependence analysis approaches, widely-used reverse engineering frameworks, and even a source-level pointer analysis tool that has access to much richer program properties. 
We also show that \sys generalizes to unseen binaries, optimizations, and obfuscations, and is drastically faster than existing approaches.
We perform extensive ablation studies to justify our design choices over other alternatives studied in previous works~\cite{pei2020trex,pei2021stateformer}.
Moreover, \sys is surprisingly accurate at many additional security-critical reverse engineering tasks, which either support or benefit from inferring memory dependencies, such as predicting memory-access regions, function signatures, and indirect procedure calls -- \sys also outperforms the state-of-the-arts on all these tasks.
% We release our code and data at \url{github.com/peikexin9/neudep}.

We make the following contributions:
\begin{enumerate}[leftmargin=*]
    \item We propose a new neural architecture that can jointly learn memory value flows from code and the corresponding traces for predicting binary memory dependencies.
    
    \item We implement our approach in \sys that contains a dedicated fusion module for learning encodings of memory addresses/trace values, and a composition learning strategy. 
    
    \item Our experimental results demonstrate that \sys is (3.5$\times$) faster and more accurate (1.5$\times$) than the state-of-the-art. 
    
    \item Our extensive ablation studies and analysis on downstream tasks suggest that our pretraining substantially improves the prediction performance and helps the model to learn value flow through different instructions.
\end{enumerate}

\section{Overview}
\label{sec:overview}

\begin{comment}
In this section, we first describe a motivating example to demonstrate the weakness of existing approaches and 
how our pretraining task on learning value flows can potentially help the model to address the problem. 
Next, we discuss how we formulate the problem of binary memory dependence analysis as a finetuning task based on a pretrained model, and describe our key design choices in \sys.
Finally, we describe three downstream reverse engineering tasks, which all benefit from precise value flow analysis across memory operations, that we used to demonstrate how much the pretrained model comprehends the instruction semantics and reason about value flows. 
\end{comment}

\subsection{Motivating Example}
\label{subsec:motiv}

\label{ex:deepvsa_fail}
\begin{figure}[!t]
\centering
\includegraphics[width=\linewidth]{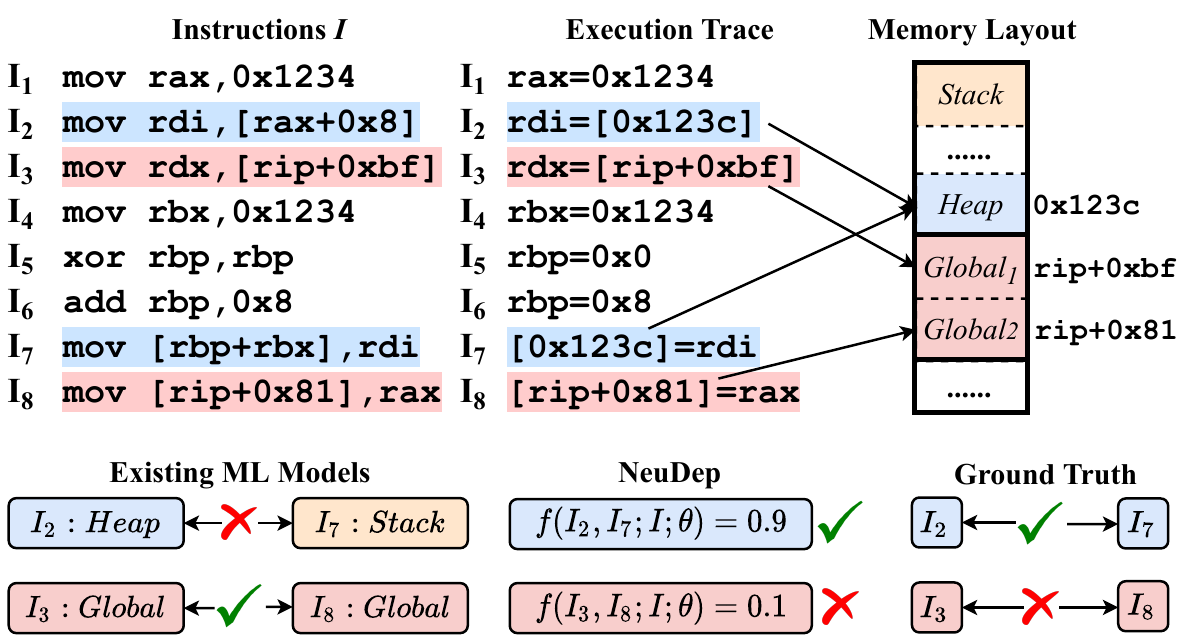}

\caption{Motivating example of predicting memory dependencies in the function \texttt{ngx\_init\_setproctitle} from \texttt{nginx-1.21.1}. 
$I_2$ and $I_7$ access the same heap variable (in \colorbox{lightblue}{blue}); $I_3$ and $I_8$ access the two \emph{different} global variables (in \colorbox{lightred}{red}). 
We find that existing ML-based approaches (learned on static code only) fail to detect $I_2$ and $I_7$ are dependent by predicting they access to different memory regions and cannot distinguish the memory accessed by $I_3$ and $I_8$ due to the prediction granularity.}
\label{fig:task_definition}
\end{figure}

Figure~\ref{fig:task_definition} shows that two instructions $I_2$: \code{mov rdi,[rax+0x8]} and $I_7$: \code{mov [rbp+rbx],rdi} access the same memory location (and are thus memory-dependent) but via different addressing registers.
To detect the dependency, the model needs to first understand that the behavior of \code{mov}: both line 1 ($I_1$) and line 4 ($I_4$) set \code{rax} and \code{rbx} to the same value.
It then needs to understand \code{xor} in line 5 sets \code{rbp} to 0, and \code{add} in line 6 performs addition and sets \code{rbp} to \texttt{0x8}. 
Finally, the model needs to compose these facts and concludes that \code{rax+0x8} is semantically equivalent to \code{rbp+rbx} in such a context, \ie they both evaluate to \texttt{0x123c}.

\para{Gap in Existing Solutions.} We find that when running the ML model trained only on static code for this task~\cite{guo2019vsa}, it mispredicts that $I_2$ and $I_7$ are not dependent as their memory-access regions ($I_2$ accesses heap while $I_7$ is mispredicted to access stack) do not intersect, possibly because its inference depends on the spurious pattern that the stack base pointer \code{rbp} is used at line 7.
Such mispredictions~\cite{guo2019vsa} might lead to a false negative by flagging two instructions as accessing non-overlapping memory regions. 

\para{Proposed Solution.} The above observation underscores the importance of encoding the knowledge about each instruction's contribution to value flows through memory and their compositions as part of the ML model. 
However, integrating a memory model as part of the encoded knowledge is challenging due to the presence of potentially complex flows involving indirect address references and their compositions. We address these challenges by designing

\begin{enumerate}[leftmargin=*]
    \item A novel training objectives to distinguish between unique memory addresses (\S\ref{subsec:problem_formulation})
    \item A dedicated fusion module sepcialized to capture the interaction between instruction, trace, and memory addresses (\S\ref{subsubsec:repr}). Our new tracing and sampling strategies (\S\ref{subsubsec:trace}) help the ML model to learn value flows across memory addresses. 
    \item Curriculum learning~\cite{bengio2009curriculum} in the training process to incrementally learn the compositional effects (\S\ref{subsubsec:comp}).
\end{enumerate}

Table~\ref{tab:example} shows some examples of how the pretraining task works and how it teaches the model to reason about value flows.

\subsection{Problem Formulation}
\label{subsec:problem_formulation}

Let $f$ denote an ML model parameterized by $\theta$.
Before directly training $f$ towards predicting memory dependencies, we pretrain $f$ to reason about the value flows (see \S\ref{subsec:pretraining_methodology} for details). 
Now consider $f$ with pretrained parameters $\theta$, we formalize the task of analyzing memory dependencies as follows.

\begin{table*}[!t]

\footnotesize
\setlength{\tabcolsep}{3pt}
\centering
\renewcommand{\arraystretch}{1.2}
\setlength\aboverulesep{0.9pt}
\setlength\belowrulesep{0.9pt}
\caption{Examples of masking (in \mask{grey}) instructions and traces (represented as input (In) and output (Out) of each instruction). The model has to dereference the memory content and interpret or synthesize instruction(s) to infer the masked parts. We include the actual operations performed by the instructions (noted in \textcolor{codegreen}{green}) and the formal semantics that the model essentially needs to learn for each example (last column).}
% \bray{This caption needs to be updated. I think the task column can go away.}
\label{tab:example}

\begin{tabular}{m{1.3\columnwidth}|cc|cc|c}
\toprule[1.1pt]
\multirow{2}{*}{Example Descriptions} & \multicolumn{2}{c|}{Instruction(s)} & \multicolumn{2}{c|}{Trace} & {Underlying} \\
& Mnemonic & Operand & In & Out & {Semantics} \\ \midrule[.9pt]

\textbf{Example 1: {\em Interpreting} memory operands in bitwise operations} & \multicolumn{2}{c|}{\comments{\textbf{rax=rax$\oplus$[rbx]}}} &  &  & \multirow{2}{*}{\begin{tabular}[c]{@{}c@{}}\inferrule{v_1=\texttt{0x2}, v_2=\texttt{0x7}\\\\ v=v_1\oplus v_2}{v=\texttt{0x5}}\end{tabular}} \\
Let the output value of \code{rax} in \code{xor rax,[rbx]} be masked.
To predict the masked value (\ie \code{rax}=\texttt{0x5}), the model needs to understand the semantics of \code{xor} on its inputs \code{rax}=\texttt{0x2} and \code{[rbx]}=\texttt{0x7}. & \code{xor} & \begin{tabular}[c]{@{}c@{}}\code{rax}\\ \code{[rbx]}\end{tabular} & \begin{tabular}[c]{@{}c@{}}\texttt{0x2}\\ \texttt{0x7}\end{tabular} & \begin{tabular}[c]{@{}c@{}}\mask{0x5}\\ \texttt{0x7}\end{tabular} &  \\ \midrule[.9pt]

\textbf{Example 2: {\em Synthesizing} Arithmetic Operations with memory operands} & \multicolumn{2}{c|}{\comments{\textbf{rbp=rbp+[rdi]}}} &  &  & \multirow{2}{*}{\begin{tabular}[c]{@{}c@{}}\inferrule{v_1=$\texttt{0x4}$, v_2=$\texttt{0x8}$\\\\ v=v_1\Diamond v_2, v=$\texttt{0xc}$}{\Diamond=\code{add}}\end{tabular}} \\
 
Let \code{add} be masked in \code{add rbp,[rdi]}. To predict the masked \code{add} (\eg out of \code{sub}, \code{mov}, etc.), the model needs to associate \code{add} to the behavior that increments the its first operand by that of its second operand.
& \mask{add} & \begin{tabular}[c]{@{}c@{}}\code{rbp}\\ \code{[rdi]}\end{tabular} & \begin{tabular}[c]{@{}c@{}}\texttt{0x4}\\ \texttt{0x8}\end{tabular} & \begin{tabular}[c]{@{}c@{}}\texttt{0xc}\\ \texttt{0x8}\end{tabular} &  \\ \midrule[.9pt]

\textbf{Example 3: Reverse {\em Interpreting} Arithmetic Operations with memory operands} & \multicolumn{2}{c|}{\comments{\textbf{rcx=rcx-[rdx]}}} &  &  & \multirow{2}{*}{\begin{tabular}[c]{@{}c@{}}\inferrule{v_2=\texttt{0x1}, v=\texttt{0x8}\\\\ v=v_1-v_2}{v_1=\texttt{0x9}}\end{tabular}} \\
 
Let the input value of \code{rcx} in \code{sub rcx,[rdx]} be masked.
To predict the masked value, the model needs to interpret \code{sub} \emph{backward} given its output \texttt{0x8} and the value of its second operand \texttt{0x1} stored on memory. 
%Such a task benefits the model when it needs to back-propagate values, and trace reverse value flows to infer memory dependencies.
& \code{sub} & \begin{tabular}[c]{@{}c@{}}\code{rcx}\\ \code{[rdx]}\end{tabular} & \begin{tabular}[c]{@{}c@{}}\mask{0x9}\\ \texttt{0x1}\end{tabular} & \begin{tabular}[c]{@{}c@{}}\texttt{0x8}\\ \texttt{0x1}\end{tabular} &  \\ \midrule[.9pt]

\textbf{Example 4: {\em Interpreting Compositions} of Multiple Memory Operations} & \multicolumn{2}{c|}{\begin{tabular}[c]{@{}c@{}}\comments{\textbf{rsp=rsp-0x8}}\\ \comments{\textbf{[rsp]=rdi}}\\ \comments{\textbf{rsi=[rsp]}}\end{tabular}} &  &  &   \\
More than one instructions are executing. 
Let the output value of \code{rsi} in \code{push rdi;mov rsi,[rsp]} be masked.
To predict the masked value, the model needs to first interpret \code{push} and understand its side effect: decrements the stack pointer \code{rsp} by 8 bytes and store the value of \code{rdi} (\texttt{0x6}) on stack referenced by \code{rsp}. & \code{push} & \begin{tabular}[c]{@{}c@{}}\code{rdi}\\ \code{rsp}\end{tabular} & \begin{tabular}[c]{@{}c@{}}\texttt{0x6}\\ \texttt{0x8}\end{tabular} & \begin{tabular}[c]{@{}c@{}}\texttt{0x6}\\ \texttt{0x0}\end{tabular} &  \multirow{2}{*}{\begin{tabular}[c]{@{}c@{}}\inferrule{v_1=v_1-\texttt{0x8}\\\\ [v_1]=\texttt{0x6}\\\\ v_2=[v_1]}{v_2=\texttt{0x6}}\end{tabular}} \\
The model then needs to first dereference the indirect addressing of \code{[rsp]} to infer \texttt{0x6} is stored at \code{rsp} (\texttt{0x0}).
It then needs to interpret \code{mov} and understand it assigns the value from its second operand to the first operand to infer the masked value to be \texttt{0x6}. & \code{mov} & \begin{tabular}[c]{@{}c@{}}\code{rsi}\\ \code{{[}rsp{]}}\end{tabular} & \begin{tabular}[c]{@{}c@{}}\texttt{0x0}\\ \texttt{0x0}\end{tabular} & \begin{tabular}[c]{@{}c@{}}\mask{0x6}\\ \texttt{0x0}\end{tabular} &  \\ \bottomrule[1.1pt]
\end{tabular}
\end{table*}

\begin{definition}[Memory Dependency Prediction]
\label{def:memory_dependency_prediction}
Given a pair of assembly code instructions $\{I_i, I_j\}$ within a code block $I$ consisting of $n$ assembly instructions: $(I_1,...,I_n)$, our neural memory dependency predictor $f$, parameterized by the pretrained weight $\theta$, predicts whether the instruction pair can access the same memory location, \ie $y=f(I_i, I_j, I;\theta), y\in[0,1]$. 
$y=0$ denotes $\{I_i, I_j\}$ do not have memory dependency, and $y=1$ denotes dependent.
\end{definition}

Any $y$ between 0 and 1 denotes the probability of $I_i$ and $I_j$ being dependent.
Figure~\ref{fig:task_definition} shows an example how our model predicts different instruction pairs. 
We elaborate on how our neural architecture implements $f$ in the above definition in \S\ref{subsec:finetune_tasks}.

\subsection{\sys's Design}
\label{subsec:overview_design}

Training the model to learn value flows for memory dependence analysis opens up several interesting design spaces, ranging from tracing to introduce diverse behaviors to designing appropriate inductive biases in the model architecture and training strategies.
We overview our design in the following and provide detailed descriptions in \S\ref{sec:method}.

\subsubsection{Trace Collection.}
\label{subsubsec:trace}

Pretraining requires high-quality training data to expose \emph{diverse} program execution.
We implement a forced execution engine~\cite{peng2014x, godefroid2014micro} to execute individual functions with full path coverage without the reliance on program test cases.
Our execution engine differs from the existing works in two key aspects.

First, we note that existing works~\cite{pei2020trex, peng2014x, egele2014blanket} implement the forced execution by violating the control flow semantics, \ie stepping through control transfer instructions, to obtain traces with high coverage. 
However, this introduces noisy traces as they are not realizable in practice.
On the contrary, respecting control transfers~\cite{pei2021stateformer} will inevitably suffer from the coverage problem, as they have to find test cases to cover different paths~\cite{brunetto2021introducing,david2016binsec,nguyen2020binary}.
To circumvent this problem, we implement a \emph{coverage-guided semantic-preserving} branch-flipping mechanism to expose diverse paths within the function without breaking the branching instructions' semantics (\S\ref{subsec:trace_methodology}).

Second, existing works do not trace behaviors of the external procedure calls~\cite{pei2020trex, pei2021stateformer}, but this is especially important to model memory operations as heap allocation is often performed via library calls (\eg via \texttt{malloc}).
Our tracing engine provides complete environment support by pre-loading the whole program and its dependent libraries.
The side effect of all function call instructions can thus be traced and logged as their input-output behavior (\S\ref{subsec:trace_methodology}).

\subsubsection{Representing and Fusing Code, Trace, and Memory.}
\label{subsubsec:repr}

Assembly instructions and their traces are highly heterogeneous, \ie instruction consists of discrete tokens like mnemonics and operands, while trace consists of mostly continuous values.
To better encode the nature of each sequence, we employ two distinct modules to learn on these two inputs and then fuse them to make the joint inference.
Specifically, we learn the instruction sequence with self-attention layers~\cite{vaswani2017attention} to encode the instructions grounded on their neighboring context.
We learn the trace values by a per-byte convolution network.
After learning a basic representation of the code and trace values, we employ a fusion module (\S\ref{subsec:arch}) to augment the contextualized instruction embeddings with the trace value embeddings.

We represent the code address space during execution as an additional input aligned to the instructions.
This helps the model stay aware of the instructions' order to their execution effect.
Moreover, we observe that \code{rip}-relative addressing is frequently used to access global variables in position-independent code. 
Therefore, feeding addresses can help the model to learn the semantics of memory addressing. 
For example, consider the following instructions from the function \texttt{quotearg\_free} in \texttt{runcon} from Coreutils-8.30.
\begin{flushleft}
\footnotesize
\setlength{\tabcolsep}{6pt}
\renewcommand{\arraystretch}{.9}

\begin{tabular}{lll}
\toprule[1.1pt]
\texttt{0x449c:} & \code{cmp [rip+0x4d65],2} & \comments{\textbf{\# rip+0x4d65=0x9208}} \\
\multicolumn{2}{l}{......} & \\
\texttt{0x44bc:} &  \code{movsxd rax,[rip+0x4d45]} & \comments{\textbf{\# rip+0x4d45=0x9208}} \\
\multicolumn{2}{l}{......} & \\
\texttt{0x450e:} &  \code{mov [rip+0x4cf0],1} & \comments{\textbf{\# rip+0x4cf0=0x9208}} \\ \bottomrule[1.1pt]
\end{tabular}
\end{flushleft}
Three instructions use \code{rip} with different offsets to access the same global variable stored at \texttt{0x9208}.
By encoding the address of each instruction, we help the model infer the value of \code{rip} and thus assist reasoning of the memory dependencies.

\subsubsection{Training Design (Composition Learning).}
\label{subsubsec:comp}

Inspired by how humans learn, we aim to develop a strategy that trains the model to gradually build up its knowledge.
Ideally, the model should start by learning easy samples and then generalize its learned knowledge by getting exposed to more challenging training samples. 
As demonstrated in Table~\ref{tab:example}, the training samples with more instructions are more challenging to predict than those with fewer instructions, as the model has to learn the \emph{compositional} execution effect of multiple instructions.
Moreover, the more masks applied, the less context the model can leverage to make the prediction, thus increasing difficulty.
Therefore, we develop a curriculum learning strategy~\cite{bengio2009curriculum} by sorting the training samples based on their length and increasing the masking rate at each training epoch.
As a result, the model always starts learning from short code pieces with fewer masks at the early batches within each epoch, and the length of the code piece and the number of masks applied are increased in later epochs.

\subsection{Additional Reverse Engineering Tasks}
\label{subsec:probing_tasks_overview}

To explore how exactly pretraining helps analyze memory dependencies, we investigate what knowledge or properties of programs the pretrained model learns.
We resort to \emph{probing}, which uses the encoded instruction representations of the pretrained model and finetunes them on the probing tasks, usually with a small number of labeled data and training epochs~\cite{manning2020emergent}.
Specifically, we consider three critical reverse engineering tasks, which either assist or benefit from analyzing memory dependencies.
If the pretrained model performs well on these reverse engineering (\ie probing) tasks, it gives evidence that pretraining has encoded useful representation for analyzing memory dependencies. 

\para{Inferring Memory Regions.}
Inferring memory-access regions helps reduce the spurious dependencies reported by VSA (\S\ref{sec:intro}).
We consider the task sketched in DeepVSA~\cite{guo2019vsa}, where the model needs to statically predict the memory region accessed by each instruction that operates on memory.

\begin{definition}[Memory Region Prediction]
\label{def:memory_region_prediction}
Given a code block consisting of a sequence of $n$ assembly instructions: $I=(I_1,...,I_n)$, a memory region predictor $f_{r}$, parameterized by the pretrained weights $\theta$, predicts the memory region accessed by each instruction: $y=f_{r}(I;\theta), y\in \mathbb{M}^n, \mathbb{M}=\{\texttt{stack,heap,global,other}\}$. 
\end{definition}

\para{Inferring Function Signature.}
Traditionally, function signatures are predicted by analyzing the memory access patterns of variables and propagating the types implied by the inferred patterns up to the function argument.
Memory dependencies help the propagating types along the dependent instructions~\cite{lee2011tie}.
The inferred variable types, in turn, also help reduce the spurious bogus dependencies, \ie two memory accesses with different types are not dependent.
We consider the task described in EKLAVYA~\cite{chua2017neural}, where the model statically predicts the function signature, including the (i) argument arity, (ii) argument types, and (iii) function return types.

\begin{definition}[Function Signature Prediction]
\label{def:function_signature_prediction}
Given an $n$-instruction procedure $P$: $P=(I_1,...,I_n)$, a function signature predictor $f_{s}$, parameterized by the pretrained weights $\theta$, predicts the function signature as follows.
(i) When $P$ is treated as callee, $f_{s}$ predicts $P$'s signature: $y=f_{s}(P;\theta)$. 
(ii) When $P$ is treated as caller, $f_{s}$ takes call site $I_c\in P$ as an additional input, and predicts the signature of the procedure that $I_c$ calls: $y=f_{s}(P, I_c;\theta)$.
In both cases, $y=(a, A, r)$ is a tuple where (i) $a\in[0,7]$ denotes argument arity with at most 7 arguments.
(ii) $A=(A_1,A_2,A_3)$ denotes $P$'s first 3 argument types: $A_i\in\{\texttt{int}, \texttt{char}, \texttt{float}, \texttt{ptr}, \texttt{enum}, \texttt{union}, \texttt{struct}\}$.
(iii) $r\in\{\texttt{int}, \texttt{char}, \texttt{float}, \texttt{ptr}, \texttt{enum}, \texttt{union}, \texttt{struct}, \texttt{void}\}$ is the procedure $P$'s return type.
\end{definition}

\para{Matching Indirect Calls.}
Analysis of memory dependencies has been extensively applied to infer indirect calls~\cite{zhang2019bda, kim2021refining}.
Therefore, we study how the pretrained model performs on this task.

\begin{definition}[Matching Indirect Calls]
\label{def:indirect_call_prediction}
Given a pair of procedures $P_i, P_j$, an indirect call predictor $f_{c}$ predicts whether $P_i$ can call $P_j$ during runtime: $y=f_{c}(P_i,P_j)$, where $y\in \{0,1\}$; $y=1$ denotes $P_i$ can call $P_j$ while $y=0$ denotes $P_i$ cannot. 
\end{definition}
Unlike the first two tasks, we define $f_{c}$ as deterministic function that takes as input the inferred function signatures (Definition~\ref{def:function_signature_prediction}) of $P_j$ and the call-site within $P_i$. 
$f_{c}$ outputs 1 if and only if the signature of $P_j$ closely matches at least one call-site signature within $P_i$. 
We elaborate on the matching criteria in \S\ref{subsec:probing_methodology}.

\section{Methodology}
\label{sec:method}

This section elaborates on the design of \sys, including the tracing framework, the model's input representation, the neural architecture, and the training tasks.

\subsection{Tracing Framework}
\label{subsec:trace_methodology}

Algorithm~\ref{alg:cov_forced_exec} shows how our tracing framework works on a procedure.
We consider the following two key designs (\S\ref{subsec:overview_design}).

\para{Environment Support.} 
As shown in Algorithm~\ref{alg:cov_forced_exec} line $1$ and $2$,
we first load the entire binary into an emulator and make a snapshot of the process image after initializing all dependent libraries. 
We then iterate every function inside the binary and execute each function (line $4$ and $5$). 
Before the execution, we restore the process memory using the saved snapshot (line $6$) to ensure that all the functions, including external library functions, are properly resolved. 

\para{Branch Flipping.} Inspired by coverage-guided fuzzing, we design a dynamic branch flipping mechanism for recording complete and diverse execution behaviors.
We first maintain a list of covered basic blocks during past execution (line 8-15). 
We then hook every conditional branch during forced execution and monitor the jump target. 
If the jump target has already been covered before and another is not covered yet, we flip the branch. 
In order to ensure the flipped branch does not introduce violations of instruction semantics, we implement a semantic-preserving mechanism by patching branch instructions with reverse conditions if it is flipped.
For example, a flipped branch instruction \code{je 0x8a} will be patched to \code{jne 0x8a}.

\begin{algorithm}[!t]
\footnotesize
\caption{Coverage-Guided Semantic-Preserving Execution} 
\label{alg:cov_forced_exec} 
\lstset{basicstyle=\ttfamily\footnotesize, breaklines=true}
\begin{spacing}{1.5}
\begin{algorithmic}[1]
\State {\textsf{Load}$(binary)$} \Comment{\textcolor{codegreen}{Load binary into emulator}}
\State {$mem = $ \textsf{Snapshot()}} \Comment{\textcolor{codegreen}{Save memory snapshot after initialization}}
\State {$covered\_bb = \{\}$}
\For{$func \in binary$}  \Comment{\textcolor{codegreen}{Loop every function}}
    \State {\textsf{Restore}$(mem)$} \Comment{\textcolor{codegreen}{Restore memory snapshot}}
    \State {\textsf{Initialize}$(stack, regs)$} \Comment{\textcolor{codegreen}{Initialize stack and registers with random values}}
    \State {\textcolor{codegreen}{/* Loop every conditional branch */}}
    \For{$cond\_branch \in $ \textsf{ForceExec}$(func)$}
        \State {\textcolor{codegreen}{/* $bb1$, $bb2$ are jump targets, $bb1$ is the default one */}}
        %\State {$bb1, bb2 =$ get\_targets$(cond\_branch)$}
        \If{ $bb1 \in covered\_bb$ \textbf{and} $bb2 \not\in covered\_bb$}
            \State {\textsf{FlipBranch}$(cond\_branch)$}
            \State {$covered\_bb$.add$(bb2)$}
        \Else
            \State {$covered\_bb$.add$(bb1)$}
        \EndIf
    \EndFor
\EndFor
\end{algorithmic}
\end{spacing}
\end{algorithm}

\subsection{Input Representation}
\label{subsec:inputs}

At a high level, \sys takes three sequences as input, \ie assembly instructions, trace values, and instruction addresses.

\para{Assembly.}
We represent the assembly instructions $I=(I_1,...,I_n)$ as $n$ ordered tuples.
Each tuple $I_i$ consists of 3 members: $I_i=(c_i, p_i, m_i)$, where $c_i$, $p_i$, $m_i$ indicate code token, position, and whether $c_i$ accesses memory, respectively.
Specifically, $c_i$ denotes the tokens obtained from tokenizing the assembly instructions, removing punctuations, and transforming all constants to \code{const}.
As we flatten each instruction to multiple tokens, we use $p_i$ to annotate the relative position of $c_i$ within the instruction to specify the instruction boundary.
Moreover, $p_i$ helps the self-attention layers, which are permutation-invariant to the input tokens, to understand the relative order of the operands.
Finally, $m_i\in\{F,T\}$ denotes whether $c_i$ accesses memory.

\begin{example}
\label{ex:assembly_code}
Consider the instruction sequence \code{add rax,0x8;mov [rax],rbx}. It will be represented as:

\begin{center}
\footnotesize
\setlength{\tabcolsep}{7pt}
\renewcommand{\arraystretch}{1.1}
\setlength\aboverulesep{0.8pt}
\setlength\belowrulesep{1.1pt}

\begin{tabular}{c|cccccc}
\toprule[1.1pt]
$I$ & $I_1$ & $I_2$ & $I_3$ & $I_4$ & $I_5$ & $I_6$ \\ \midrule[.9pt]
\multirow{3}{*}{$\left(\begin{tabular}[c]{@{}c@{}}$c$\\ $p$\\ $m$\end{tabular}\right)$} & \multirow{3}{*}{$\left(\begin{tabular}[c]{@{}c@{}}\code{add}\\ $1$\\ F\end{tabular}\right)$} & \multirow{3}{*}{$\left(\begin{tabular}[c]{@{}c@{}}\code{rax}\\ $2$\\ F\end{tabular}\right)$} & \multirow{3}{*}{$\left(\begin{tabular}[c]{@{}c@{}}\code{const}\\ $3$\\ F\end{tabular}\right)$} & \multirow{3}{*}{$\left(\begin{tabular}[c]{@{}c@{}}\code{mov}\\ $1$\\ F\end{tabular}\right)$} & \multirow{3}{*}{$\left(\begin{tabular}[c]{@{}c@{}}\code{rax}\\ $2$\\ T\end{tabular}\right)$} & \multirow{3}{*}{$\left(\begin{tabular}[c]{@{}c@{}}\code{rbx}\\ $3$\\ F\end{tabular}\right)$} \\
 &  &  &  &  &  &  \\
 &  &  &  &  &  &  \\ \bottomrule[1.1pt]
\end{tabular}
\end{center}
\end{example}

\para{Trace.}
We represent the trace values using $T=(T_1, ..., T_n)$ aligned to the assembly instruction sequence $I$.
Each $T_i\in T$ consists of a list $T_i=(b^i_1,...,b^i_8)$, where the numeric value is a (padded) 8-byte values $b^i_{j\in[1,8]}$.
This reduces a prohibitively large vocabulary ($2^{64}$) to a much more manageable size ($256$)~\cite{pei2021stateformer}.
The most and least significant byte is $b_1$ and $b_8$, respectively.
We further normalize each byte $b^i_{j\in[1,8]}$ into $[0,1)$ to stablize the training.
% by dividing each byte by 256 to further stablize the training, \eg $\texttt{0x80}/256=0.5$.
For instruction tuples $I_i$ whose $c_i$ is not a register or a constant, its aligned trace values $T_i$ contains 8 dummy values (\texttt{-}), which will not be masked during pretraining (\S\ref{subsec:pretraining_methodology}).
For $T_i$ whose aligned $I_i$ is not executed, we assign the value $b_j=$\texttt{0x100}$, \forall j\in[1,8]$.
In pretraining, to predict the trace value consisting of all 100s instead of regular bytes, the model needs to determine whether the corresponding assembly instructions are executed by reasoning the branch predicate and control flow.

\begin{example}
Consider the following 4 instructions: \code{add rax,0x8;cmp rax,0x10;je 0x1004a8b5f;push rdi}; input \code{rax}=\texttt{0x0}. $T$ will look like (aligned with $c_i$):

\begin{center}
\centering
\footnotesize
\setlength{\tabcolsep}{2.5pt}
\renewcommand{\arraystretch}{1.1}
\setlength\aboverulesep{0.8pt}
\setlength\belowrulesep{0.8pt}
\begin{tabular}{c|cccccccccc}
\toprule[1.1pt]
$c$ & \code{add} & \code{rax} & \code{const} & \code{cmp} & \code{rax} & \code{const} & \code{je} & \code{const} & \code{push} & \code{rdi} \\ \midrule[.9pt]
$T$ & $T_1$ & $T_2$ & $T_3$ & $T_4$ & $T_5$ & $T_6$ & $T_7$ & $T_8$ & $T_9$ & $T_{10}^*$ \\ \midrule[.9pt]
\multirow{8}{*}{$\left(\begin{tabular}[c]{@{}c@{}}$b_1$\\ $b_2$\\ $b_3$\\ $b_4$\\ $b_5$\\ $b_6$\\ $b_7$\\ $b_8$ \end{tabular}\right)$} & \multirow{8}{*}{$\left(\begin{tabular}[c]{@{}c@{}}\texttt{-}\\ \texttt{-}\\ \texttt{-}\\ \texttt{-}\\ \texttt{-}\\ \texttt{-}\\ \texttt{-}\\ \texttt{-} \end{tabular}\right)$} & \multirow{8}{*}{$\left(\begin{tabular}[c]{@{}c@{}}\texttt{00}\\ \texttt{00}\\ \texttt{00}\\ \texttt{00}\\ \texttt{00}\\ \texttt{00}\\ \texttt{00}\\ \texttt{00} \end{tabular}\right)$} & \multirow{8}{*}{$\left(\begin{tabular}[c]{@{}c@{}}\texttt{00}\\ \texttt{00}\\ \texttt{00}\\ \texttt{00}\\ \texttt{00}\\ \texttt{00}\\ \texttt{00}\\ \texttt{08} \end{tabular}\right)$} & \multirow{8}{*}{$\left(\begin{tabular}[c]{@{}c@{}}\texttt{-}\\ \texttt{-}\\ \texttt{-}\\ \texttt{-}\\ \texttt{-}\\ \texttt{-}\\ \texttt{-}\\ \texttt{-} \end{tabular}\right)$} & \multirow{8}{*}{$\left(\begin{tabular}[c]{@{}c@{}}\texttt{00}\\ \texttt{00}\\ \texttt{00}\\ \texttt{00}\\ \texttt{00}\\ \texttt{00}\\ \texttt{00}\\ \texttt{08} \end{tabular}\right)$} & \multirow{8}{*}{$\left(\begin{tabular}[c]{@{}c@{}}\texttt{00}\\ \texttt{00}\\ \texttt{00}\\ \texttt{00}\\ \texttt{00}\\ \texttt{00}\\ \texttt{00}\\ \texttt{10} \end{tabular}\right)$} & \multirow{8}{*}{$\left(\begin{tabular}[c]{@{}c@{}}\texttt{-}\\ \texttt{-}\\ \texttt{-}\\ \texttt{-}\\ \texttt{-}\\ \texttt{-}\\ \texttt{-}\\ \texttt{-} \end{tabular}\right)$} & \multirow{8}{*}{$\left(\begin{tabular}[c]{@{}c@{}}\texttt{00}\\ \texttt{00}\\ \texttt{00}\\ \texttt{01}\\ \texttt{00}\\ \texttt{4a}\\ \texttt{8b}\\ \texttt{5f} \end{tabular}\right)$} & \multirow{8}{*}{$\left(\begin{tabular}[c]{@{}c@{}}\texttt{-}\\ \texttt{-}\\ \texttt{-}\\ \texttt{-}\\ \texttt{-}\\ \texttt{-}\\ \texttt{-}\\ \texttt{-} \end{tabular}\right)$} & \multirow{8}{*}{$\left(\begin{tabular}[c]{@{}c@{}}\texttt{100$^*$}\\ \texttt{100$^*$}\\ \texttt{100$^*$}\\ \texttt{100$^*$}\\ \texttt{100$^*$}\\ \texttt{100$^*$}\\ \texttt{100$^*$}\\ \texttt{100$^*$} \end{tabular}\right)$} \\
 &  &  &  &  &  &  &  &  &  & \\
 &  &  &  &  &  &  &  &  &  & \\
 &  &  &  &  &  &  &  &  &  & \\
 &  &  &  &  &  &  &  &  &  & \\
 &  &  &  &  &  &  &  &  &  & \\
 &  &  &  &  &  &  &  &  &  & \\
 &  &  &  &  &  &  &  &  &  & \\ \bottomrule[1.1pt]
\multicolumn{11}{l}{\multirow{2}{*}{\begin{tabular}[c]{@{}l@{}}\scriptsize *We show the byte value before normalization to save space. As $T_{10}$ corresponds to \code{rdi}, which\\ \scriptsize is not executed, its value is 1, which is \texttt{0x100} before normalizing. \end{tabular}}} \\
\multicolumn{11}{l}{}
\end{tabular}
\end{center}

\end{example}

\para{Address.}
Similar to trace value sequence $T$, we represent the address of each instruction (when loaded in memory) as $n$ ordered lists, $A=(A_1,..., A_n)$, aligned to $I$.
$A_i\in A$ consists of 8 bytes organized as an ordered list: $A_i=(b^i_1, b^i_2,...,b^i_8)$.
On a 64-bit architecture, 8 bytes are enough to represent all possible virtual addresses of a running program. 
For $c_i$ within one instruction (\eg Example~\ref{ex:assembly_code}), they share the same instruction address.

\begin{example}
Consider 2 instructions: \code{push rbp;jmp rax} start from the address \texttt{0x14a8b}. $A$ will look like (aligned with $c_i$):

\begin{center}
\footnotesize
\setlength{\tabcolsep}{12pt}
\renewcommand{\arraystretch}{1.1}
\setlength\aboverulesep{0.8pt}
\setlength\belowrulesep{0.8pt}

\begin{tabular}{c|cccc}
\toprule[1.1pt]
$c$ & \code{push} & \code{rbp} & \code{jmp} & \code{rax} \\ \midrule[.9pt]
$A$ & $A_1$ & $A_2$ & $A_3$ & $A_4$ \\ \midrule[.9pt]
\multirow{3}{*}{$\left(\begin{tabular}[c]{@{}c@{}} $b_6$\\ $b_7$\\ $b_8$ \end{tabular}\right)$} & \multirow{3}{*}{$\left(\begin{tabular}[c]{@{}c@{}} \texttt{01}\\ \texttt{4a}\\ \texttt{8b} \end{tabular}\right)$} & \multirow{3}{*}{$\left(\begin{tabular}[c]{@{}c@{}} \texttt{01}\\ \texttt{4a}\\ \texttt{8b} \end{tabular}\right)$} & \multirow{3}{*}{$\left(\begin{tabular}[c]{@{}c@{}} \texttt{01}\\ \texttt{4a}\\ \texttt{8c} \end{tabular}\right)$} & \multirow{3}{*}{$\left(\begin{tabular}[c]{@{}c@{}} \texttt{01}\\ \texttt{4a}\\ \texttt{8c} \end{tabular}\right)$} \\
 &  &  &  & \\
 &  &  &  & \\ \bottomrule[1.1pt]
\multicolumn{5}{l}{\scriptsize We omit showing $b_1,...,b_5$ as they are all zeros} \\
\end{tabular}
\end{center}
As the machine instruction of \code{push rbp} only takes one byte (\texttt{0x55}), the addresses of two instructions are off by one byte.
% (\eg \texttt{0x14a8b} for \code{push rbp} and \texttt{0x14a8b} for \code{jmp rax}).
\end{example}

\subsection{\sys Architecture}
\label{subsec:arch}

\begin{figure}[t]
\centering
\includegraphics[width=\linewidth]{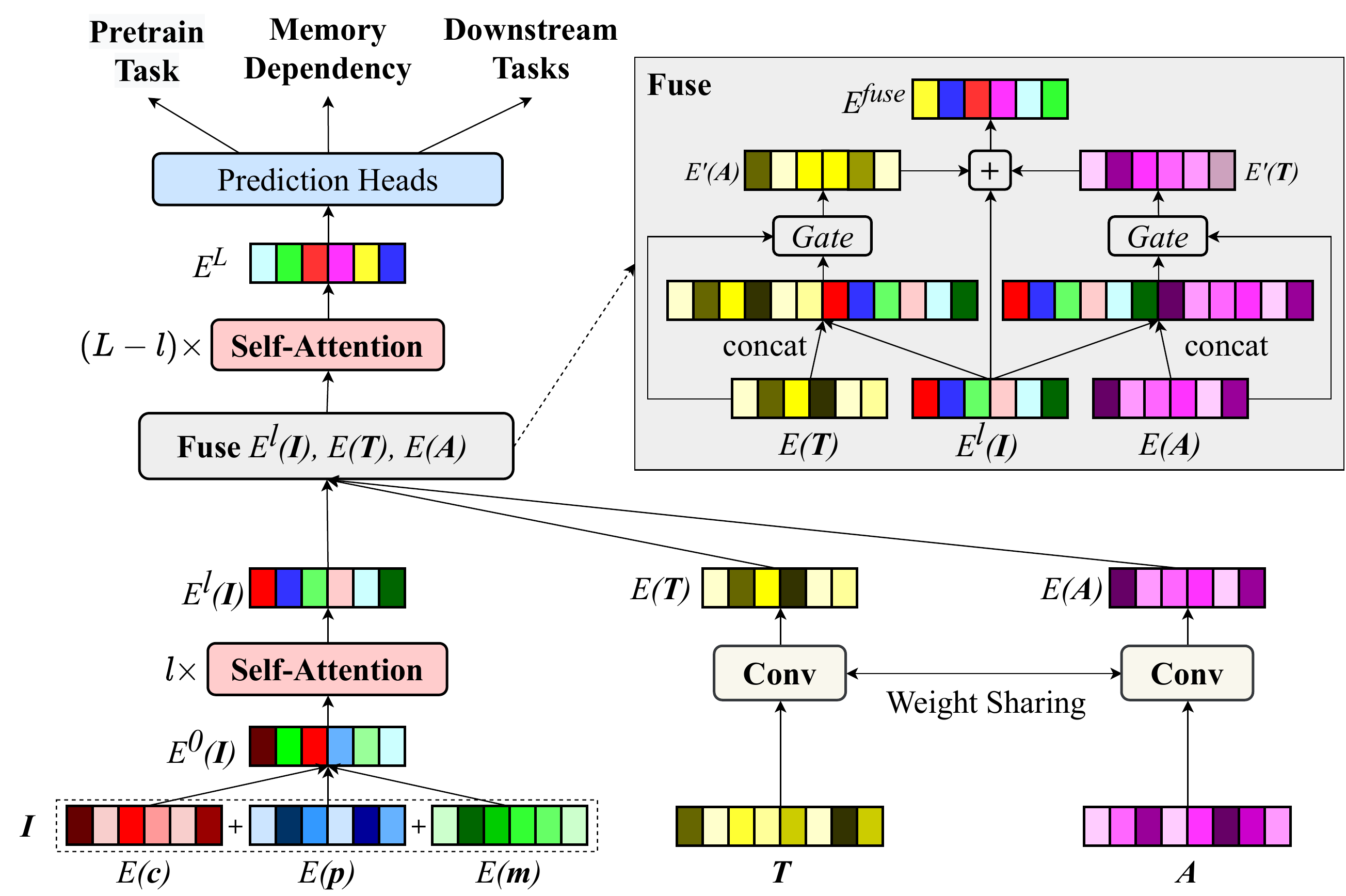}

\caption{\sys's high-level architecture with details of the fusion module. It takes as input 3 sequences: the instructions $I$, trace values $T$, and code addresses $A$ (\S\ref{subsec:inputs}), where $I$ is embedded by $l$ self-attention layers, and $T$ and $A$ are embedded by convolution networks. They are then fused by a fusion module. The fused embeddings go through another $L-l$ self-attention layers and output the final embeddings $E^L$ (\S\ref{subsec:arch}).}

\label{fig:arch}
\end{figure}

Figure~\ref{fig:arch} illustrates \sys's architecture.
In the following, we describe how the inputs (\S\ref{subsec:inputs}) are embedded, fused, and further processed to make the prediction.
All these steps are handled by neural modules that can be stacked together and trained end-to-end.

\para{Input Embeddings.} 
Let $d$ denote the embedding dimension, we denote the embeddings of each tuple $(c_i,p_i,m_i$) as $E(c_i), E(p_i), E(m_i)\in \mathds{R}^{d}$.
We sum these embeddings to form the embeddings of $I_i$: $E(I_i)=E(c_i)+E(p_i)+E(m_i)$.
We denote the embedding of all tokens as $E^0(I)=(E(I_1),...,E(I_n))$, representing the instructions' embeddings before the first self-attention layer.
We first apply $l$ self-attention layers on $E^0(I)$ to learn the contextual embeddings of the assembly instructions: $E^l(I)$.

To embed the 8-byte values in $T$ and $A$ into a space that preserves their numerical properties, we employ a convolution network with 8 kernels to learn how bytes within each neighboring size interact with each other.
Let $C_{w}$ denote applying a convolution filter with width $w$ and output channel $O_w$, we first apply 8 convolution filters on $T_i=(b_1,...,b_8)$ and concatenate them: $C_{out}=concat(\phi(max(C_1(T_i))),...,\phi(max(C_8(T_i))))$.
Here $\phi$ denotes an activation function, and we use ReLU in this paper.
$C_{out}\in\mathds{R}^{\sum^8_{w=1} O_w}$ is the concatenated result.
We then transform $C_{out}$ by a highway network~\cite{srivastava2015highway} (see Appendix) and obtain the embedding for $T_i$: $E(T_i)$.

To learn a universal value representation from 8 bytes, we share the weights of this network by applying it on both $T$ and $A$.
Therefore, $E(A_i)$ is embedded similarly as described above.

\para{Fusing Heterogenous Inputs.}
Intuitively, after embedding all the inputs, they are expected to carry basic meaning in their own modalities.
We then employ a fusion module that augments the instruction embedding by its traces and address.
Specifically, let $G_{T_i}=\sigma(MLP(concat(E^l(I_i), E(T_i))))$ and $G_{A_i}=\sigma(MLP(concat(E^l(I_i),E(A_i))))$ denote the learned gates that control how much $E(T_i)$ and $E(A_i)$ should be fused in $E^l(I_i)$, we have: 
\begin{equation*}
    E^{fuse}_i=E^l(I_i)+G_{T_i}\cdot MLP(E(T_i))+G_{A_i}\cdot MLP(E(A_i))
\end{equation*}

\para{Cross-Modality Inference.}
Let $L$ denote the total number of self-attention layers, in which $l$-th layers are used to learn the instruction representation $E^l(I)$.
We feed $E^{fuse}=(E_1^{fuse},...,E_n^{fuse})$ to the remaining $L-l$ layers.
On top of the last self-attention layer, we obtain $E^L=(E_1^L,...,E_n^L)$ and employ trainable prediction heads for pretraining (\S\ref{subsec:pretraining_methodology}), finetuning (\S\ref{subsec:finetune_tasks}), and probing (\S\ref{subsec:probing_methodology}).

\subsection{Pretraining to Interpret and Synthesize Code}
\label{subsec:pretraining_methodology}

We give the formal definition of our pretraining task as follows.

\begin{definition}[Pretraining]
\label{def:pretrain}
Given (i) a code block $I=(I_1, ..., I_n)$, (ii) its trace $T$: $T=(T_1, ..., T_n)$, and (iii) a mask rate $r$, we pretrain the model $f$, parameterized by $\theta$, by the following training objectives.
\begin{enumerate}[leftmargin=*]
    \item \emph{Interpret} $I$: predict the masked trace $T_{MT}$: $T_{MT}\subseteq T, |T_{MT}|=|T|\cdot r$, given $I$ and $T - T_{MT}$: $T_{MT}=f(I,T - T_{MT};\theta)$.
    \item \emph{Synthesize} $I$: predict the masked instructions $I_{MI}$: $I_{MI}\subseteq I, |I_{MI}|=|I|\cdot r$, given $I-I_{MI}$ and $T$: $I_{MI}=f(I-I_{MI},T ;\theta)$.
    \item \emph{Both}: predict both $I_{MI}$ and $T_{MT}$ given $I-I_{MI}$ and $T-T_{MT}$: $I_{MI},T_{MT}=f(I-I_{MI},T-T_{MT};\theta)$.
\end{enumerate}
\end{definition}

Specifically, the pretraining takes as input the output of the last self-attention layer $E^L=(E_1^L,...,E_n^L)$, and minimize the (1) cross-entropy (CE) between the predicted masked code $\hat{c}$ and the actual code $c_{MI}$, and the (2) mean squared error (MSE) between predicted masked values (8 bytes) $\hat{T}_{MT}$ and the actual values $T_{MT}$:
\begin{equation}
\label{eq:pretrain}
\centering
\argmin_{\theta} \sum_{i\in MI} -c_i\log(\hat{c}_i)+ \alpha\sum_{j\in MT}(\hat{T}_j-T_j)^2
\end{equation}
$\theta$ denote the trainable parameters \sys's model (\S\ref{subsec:arch}) and the prediction heads: $MLP_c$ and $MLP_T$, two multilayer perceptrons that take $E^L$ as input and predict the masked instructions: $\hat{c}_{MI}=MLP_c(E^L_{MI})$ and trace values: $\hat{T}_{MT}=MLP_T(E^L_{MT})$.

\para{Composition Learning.}
We increase the masking percentage $r$ (Definition~\ref{def:pretrain}) at each epoch.
Let $L$, $U$ denote the lower and upper bound of the $r$, respectively, and $EPO_{pretrain}$ denote the pretraining epochs, at $k$-th epoch, $r=L+(U-L)\times(k-1)/EPO_{pretrain}$.

\subsection{Finetuning to Predict Memory Dependencies}
\label{subsec:finetune_tasks}

As shown in Figure~\ref{fig:workflow}, after the model is pretrained, we let the model predict value flows between instructions based on its learned representation of assembly code without traces.
To this end, we detach the fusion module and the convolution module for embedding the trace $T$ and addresses $A$ (right part in Figure~\ref{fig:arch}) and directly stack the upper $L-l$ self-attentions on top of the first $l$ self-attention layers.

Given $E^L=(E^L_1,...,E^L_n)$ (\S\ref{subsec:arch}), we employ a prediction head $MLP_{dep}$ that minimizes the binary cross-entropy (BCE) between the predicted dependency $\hat{y}$ of $\{I_i, I_j\}\subseteq I$ and their ground truth $y$ (Definition~\ref{def:memory_dependency_prediction}): $\argmin_{\theta} -y\cdot \log \hat{y}-(1-y)\cdot \log (1-\hat{y})$, where 
% $\hat{y} = MLP_{dep}(concat(\psi_a(E^L), E^L_i, E^L_j, |E^L_i-E^L_j|, E^L_i\odot E^L_j))$.
\begin{equation*}
\label{eq:finetune}
\centering
    \hat{y} = MLP_{dep}(concat(\psi_a(E^L), E^L_i, E^L_j, |E^L_i-E^L_j|, E^L_i\odot E^L_j))
\end{equation*}
Here $\psi_a$ denotes taking the mean pooling of $E^L$ and $\odot$ denotes element-wise multiplication.
This results in the input shape to $MLP_{dep}$ to be $\mathds{R}^{5d}$.

\subsection{Downstream Reverse Engineering Tasks}
\label{subsec:probing_methodology}

As described in \S\ref{subsec:probing_tasks_overview}, we consider three security-critical reverse engineering tasks as our probing tasks.
We follow the similar setup in \S\ref{subsec:finetune_tasks} and stack separate prediction heads on top of $E^L$, and train with additional training samples collected for probing.

\para{Inferring Memory Regions.}
Given the output of the last self-attention layers $E^L=(E^L_1,...,E^L_n)$, we stack a prediction head $MLP_{r}$ that predicts the memory-access regions for each instruction $I_i$.
The training task then minimizes the sum of cross-entropy between the predicted memory regions $\hat{y}$ of \emph{each} instruction and their ground truth memory region $y$: $\argmin_{\theta}\sum^n_{i=1} -y_i\cdot \log (MLP_{r}(E_i^L))$.
% \begin{equation*}
% \label{eq:memory_region}
% \centering
%     \argmin_{\theta}\sum^n_{i=1} -y_i\cdot \log (MLP_{region}(E_i^L))
% \end{equation*}

\para{Inferring Function Signature.}
As shown in Definition~\ref{def:function_signature_prediction}, predicting function signatures consists of predicting 5 types of labels: $\{a, A_1, A_2, A_3, r\}$.
% Definition \ref{def:function_signature_prediction} tells us that a procedure's overall signature consists of 5 distinct components $c \in \{a$, $arg_{1-3}$, $r\}$. 
For each label, we create two prediction heads: $MLP_{caller}$ and $MLP_{callee}$.
% As a result, we create ten separate prediction heads.
For example, $MLP^{a}_{caller}$ takes as input the embedding corresponding to the call site $c\in[1,n]$ from the last self-attention layer $E^L$, and predicts the number of arguments that the call site prepares: $a=MLP^a_{caller}(E^L_c)$. 
$MLP^{a}_{callee}$ takes as input the embeddings from the last self-attention layer $E^L$ and predicts the number of arguments the callee expects: $a=MLP^a_{callee}(\psi_a(E^L))$ where $\psi_a$ denotes the average pooling of all embeddings in $E^L$.
% We treat each component $c$ separately by defining two prediction heads specific to $c$, $MLP^{(c)}_{caller}$ (which predicts the value of $c$ given a caller input) and $MLP^{(c)}_{callee}$ (which handles callees). 
% Thus, we create 10 prediction tasks in total. 
The training objective for each head then minimizes the cross-entropy loss between the predicted label and the ground truth label.

\para{Matching Indirect Calls.}
Given the signatures of a call site $P_i$ and a callee $P_j$, we implement the indirect call predictor $f_{c}$ (Definition \ref{def:indirect_call_prediction}) by considering the following four criteria.
(i) \emph{Loose arity}: $P_i$ must prepare at least as many arguments as $P_j$ accepts. 
(ii) \emph{Strict arity}: the arities of $P_i, P_j$ must match exactly.
(iii) \emph{Argument type}: the types of $P_i$'s first three arguments must match $P_j$'s argument types in at least 2 or 3 positions.
(iv) \emph{Return type}: if $P_i$ is non-void, then $P_j$ must be non-void.
The four criteria can be composed to determine whether $P_i, P_j$ matches. 
We evaluate 8 distinct compositions in \S\ref{subsec:rq3}.

\section{Implementation and Setup}
\label{sec:impl}

% In this paper, we focus on x86-64 binaries, but \sys does not have any architecture-specific design that prevents from learning on binaries running on other architectures. 
We implement \sys's tracing framework in Qiling~\cite{qiling} and the model architecture based on PyTorch.
We run all experiments and baselines on a Linux server, with Intel Xeon 4214 at 2.20GHz with 48 virtual cores, 188GB RAM, and 4 Nvidia RTX 2080Ti GPUs.

\para{Dataset.}
We collect 41 open-source projects, ranging from utility libraries like Binutils to popular libraries like OpenSSL (see Appendix).
% We compiled 13 ourselves, and used 28 different projects present from Pang~\etal\cite{pang2021sok} dataset. 
% These projects include Diffutils-3.1, OpenSSL-1.0.1u, OpenSSL1.0.1f, OpenSSL-1.1.01, BusyBox-1.32.0, Curl-7.71.1, Gmp-6.2.0, ImageMagick-7.0.10.46, libmicrohttpd-0.9.71, libtomcrypt-1.18.2, sqlite-3.35.0, zlib-1.2.11, gawk-5.1.0, putty-0.74 - NOW ONTO x86-SOK - Unzip-6.0, Coreutils-8.30, 7-zip-19, Findutils-4.4, Binutils-2.26, Tiff-4.0, Putty-0.73, D8-6.4, Filezilla-3.44.2, Busybox-1.31, Protobuf-c-1, ZSH-5.7.1, VIM-8.1, XML2-2.9.8, Openssh-8.0, Git-2.23, Lighttpd-1.4.54, Mysqld-5.7.27, Nginx-1.15.0, SQLite-3.32.0, Glibc-2.27, libpcap-1.9.0, libv8-6.4, libtiff-4.0.10, libxml2-2.9.8, libsqlite-3.32.0, libprotobuf-c-1.3.2. 
We compiled these projects with 4 optimizations, \ie O0-O3, using GCC-9.3.0, and 4 obfuscations based on Clang-8~\cite{hikari}, \ie bogus control flow (bcf), control flow flattening (cff), basic block splitting (spl), and instruction substitution (sub).
% We preprocess each procedure within all binaries into the input format described in \S\ref{subsec:inputs}.
% As a result, we do not expose \sys with any symbol or debugging information. 
% Table~\ref{tab:dataset} summarizes the statistics of the dataset.
Among the 41 projects, we select 9 projects as our finetuning set and the rest for pretraining.
They include bash-5.0, bc-1.07.1, binutis-2.30, bison-3.3.2, cflow-1.6, coreutils-8.30, curl-7.76.0, findutils-4.7.0, gawk-5.1.0.
The 9 projects have disparate functionalities and sizes such that they are diverse and representative of real-world software.
We perform static disassembly (taking less than 0.1 seconds per input) followed by a simply post-processing to parse the raw assembly into the format that the model accepts (\S\ref{subsec:inputs}).

% \begin{table}[!t]

% \footnotesize
% \setlength{\tabcolsep}{14pt}
% \centering
% \renewcommand{\arraystretch}{1}
% \setlength\aboverulesep{0.2pt}
% \setlength\belowrulesep{0.2pt}
% % \rowcolors{2}{}{lightblue}

% \caption{The statistics of our datasets, categorized by optimization (OPT) and obfuscation (OBF).
% % The functions with the same name in different version of projects or in different projects are considered as different functions. 
% }
% \label{tab:dataset}

% % \rowcolors{1}{}{lightblue}

% \begin{tabular}{rlrr}
% \toprule[1.1pt]
% Flags    &  Size (GB) & \# Inst & \# Func \\ \midrule[.9pt]
% \rowcolor{shadecolor} O0    & 3.46 & 40,679,110  &616,773 \\
%                         O1& 3.51  &  38,877,129   &  596,857 \\
% \rowcolor{shadecolor} O2    & 3.53 &  39,063,879 & 597,283 \\
%                     O3    &  3.55 &   39,852,540  &  594,918 \\ \midrule[.9pt]
% \rowcolor{shadecolor} bcf   & 0.12 &  11,065,100 & 56,873 \\
%                     cff   & 0.10  & 10,358,084 &  57,487\\
% \rowcolor{shadecolor} spl   & 0.10 & 8,301,924 & 56,873 \\ 
%                     sub   &  0.08 &  6,154,209 & 56,871\\ \midrule[.9pt]
% \rowcolor{shadecolor} Total & 14.45 & 194,351,975 & 2,633,935 \\ \bottomrule[1.1pt]
% \end{tabular}
% \end{table}

\para{Ground Truth Dependencies.}
We follow~\cite{zhang2019bda} by using dynamic analysis to collect the ground truth memory dependencies.
To quantify how \sys and the baselines perform, we measure the detected dependencies among the reference ones (\emph{detect}) and mark the rest as \emph{miss}; we treat the predicted dependencies not included in the references as potential false positives (FP).

\para{Baselines.}
We compare \sys to Angr~\cite{wang2017angr}, Ghidra~\cite{ghidra}, SVF~\cite{sui2016svf}, and DeepVSA~\cite{guo2019vsa}.
% We run each baseline on the same testing set we collected for evaluation (see above).
% We describe them in the following.
% Angr is a popular binary analysis framework implementing VSA to analyze memory dependencies.
% Angr restricts the slices on which VSA is applied with only three basic blocks, mitigating the scalability issue by trading the precision.
% Even with such a treatment, we notice that Angr times out on 71\% functions in our dataset (Table~\ref{tab:dataset}).
% Therefore, we make sure the functions from which we sample the reference instruction pair do not cause timeout for Angr.
% It analyzes value flows at the basic block level, so we treat it a true positive as long as its detected dependent basic blocks include the ground truth instruction pairs. 
% This thus \emph{over-estimates} Angr's detection precision.
% \kexin{Dongdong could you elaborate}
% Ghidra is a powerful reverse engineering framework. 
% It associates every memory accesses with one or more variables.
% If two instructions are flagged accessing same variables, we mark it as predicted dependent by Ghidra. 
% SVF is a compile-time pointer analysis framework that generates memory dependencies at LLVM IR level.
As SVF does support dumping its result to the compiled binary (confirmed with the authors)~\cite{sui2016svf}, we propagate its result using the DWARF information.
% To evaluate how SVF performs at the binary executable, we propagate its IR-level dependencies to the binary-level by mapping the IR statements to the source code statements, and then map the source statements to binary instructions based on the debugging information. 
% Even with debugging information, propagating IR-level memory dependencies to binaries can introduce nontrivial noises~\cite{guo2005practical}.
% For example, one statement can map to multiple binary instructions and some of them do not reference any memory addresses, or the IR undergoes other optimizations.
As one source statement can map to multiple assembly instructions, we treat it as a true positive if its detected dependencies include the ground truth instruction pair.
We thus omit evaluating SVF on the obfuscated binaries as the obfuscator significantly distorts the mapping in DWARF.
% This thus also \emph{over-estimates} SVF's detection precision. 
% DeepVSA predicts memory-access regions (Definition~\ref{def:memory_region_prediction}) based on a hierarchical attention neural net.
% It predicts whether two instructions are dependent based on their accessed memory regions.
For DeepVSA, its VSA implementation requires taking a crash dump as input and does not work for general memory dependence analysis.
Therefore, we run its released trained model and use its predicted memory region to determine whether two memory-access instructions are dependent.
We note that PalmTree~\cite{li2021palmtree} also compared to DeepVSA on the standalone memory region prediction task without running its VSA module. However, as PalmTree does not release its trained model for memory region prediction, we cannot run PalmTree on our dataset to predict memory dependencies.
Therefore, we instead compare \sys to PalmTree and its evaluated baselines (including DeepVSA) in our probing studies (\ref{subsec:rq3}).

BDA~\cite{zhang2019bda} is the state-of-the-art binary memory dependence analysis tool, to the best of our knowledge. 
We reached out to the authors and confirmed that BDA mainly targets reducing false negatives in the inter-procedural setting, and they evaluated it on only O0 binaries.
Per our requests, BDA authors performed preliminary studies and observed BDA achieves low miss rate (0.02\%), but suffers from high false positive rate and runtime overhead.
For example, on readelf compiled by O0, BDA has around 2.23\% precision (detecting 5,742 true dependencies out of a total of 256,596 predicted dependencies).  
Due to the different focuses between BDA and \sys, we thus omit including BDA results to avoid unfair comparison.

% We compare \sys to 9 baselines for different sub-tasks (see \S\ref{subsec:task_formulation}).
% In addition to evaluating the performance on explicit memory dependence tasks, we study how \sys's pretrained knowledge assists in performing type inference of function arguments and matching indirect function calls, two binary analysis tasks that benefit from memory dependence analysis.
For probing tasks, we compare \sys to (i) PalmTree~\cite{li2021palmtree} and the other baselines that PalmTree evaluated such as DeepVSA~\cite{guo2019vsa}, Asm2Vec~\cite{ding2019asm2vec}, and Instruction2Vec~\cite{lee2017learning}, on predicting memory-access regions, (ii) EKLAVYA~\cite{chua2017neural} on predicting function signatures, and (iii) EKLAVYA and TypeArmor~\cite{van2016tough} on predicting indirect calls.
% For inferring indirect calls, we compare to TypeArmor~\cite{van2016tough}, the state-of-the-art approach detecting indirect calls.
% TypeArmor is a binary hardening tool that uses static register-access (read before write vs write before read vs unused) analysis to infer function arity and a binary return type classification (void or non-void). 
% TypeArmor infers these values together by following forward edges in callees and backwards edges for callers.
% Particularly, EKLAVYA is an ML-based tool that trains a recurrent neural net for each function signature prediction task (\S\ref{subsec:probing_methodology}). 
% We compare \sys directly to the reported accuracies in their paper.
% To compare on tasks not reported in their paper, \eg predicting return types, we train their model using their released code. 
% TypeArmor is a binary hardening tool that analyzes register-access patterns to infer function arity and return type, which are then used to match indirect calls. 
% As we observe that TypeArmor crashes on part of our testing binaries, we restrict our test set to those that do not crash TypeArmor. 
% This thus over-estimates TypeArmor's performance.

\para{Hyperparameters.}
% In this paper, we fix both the pretraining and finetuning epochs to be 10.
For composition learning, we set $L=0.2$ and $U=0.8$ (\S\ref{subsec:pretraining_methodology}).
For pretraining (Equation~\ref{eq:pretrain}), we set $\alpha=100$ by observing that MSE loss is around 100$\times$ smaller than that of CE in our experiments.

\section{Evaluation}
\label{sec:eval}

\begin{table*}[!t]

\footnotesize
\setlength{\tabcolsep}{7pt}
\centering
\renewcommand{\arraystretch}{1.1}
\setlength\aboverulesep{0.4pt}
\setlength\belowrulesep{0.4pt}

\caption{\sys and other baselines' results on the test set categorized by the compiler optimizations and obfuscations.
}
\label{tab:result}

\begin{tabular}{rrlllllllllllllll}
\toprule[1.1pt]
\multirow{2}{*}{Flags} & \multirow{2}{*}{\# Dep} & \multicolumn{3}{c}{Angr} & \multicolumn{3}{c}{Ghidra} & \multicolumn{3}{c}{SVF} & \multicolumn{3}{c}{DeepVSA$^*$} & \multicolumn{3}{c}{\sys} \\
 &  & Detect & Miss & FP & Detect & Miss & FP & Detect & Miss & FP & Detect & Miss & FP & Detect & Miss & FP \\ \midrule[.9pt]
\rowcolor{shadecolor} O0 & 1,013  & 28 & 985 & 16 & 355 & 658 & 7  & 380 & 633 & 392 & 420 & 593 & 329 & \textbf{930} & \textbf{83} & \textbf{147} \\
                      O1 & 1,310  & 12 & 1,298 & 14 & 387 & 923 & 19 & 486 & 824 & 1,474 & 617 & 693 & 2,013 & \textbf{898} & \textbf{412} & \textbf{576} \\
\rowcolor{shadecolor} O2 & 1,103  & 14 &1,089  & 10 & 330 & 773 & 6  & 403 & 698 & 1,392 & 464 & 639 & 1,512 & \textbf{822} & \textbf{281} & \textbf{480} \\
                      O3 & 1,132  & 14 & 1,118 & 14 & 332 & 800 & 14 & 425 & 707 & 1,351 & 472 & 660 & 1,527 & \textbf{827} & \textbf{305} & \textbf{501} \\ \midrule[.9pt]
\rowcolor{shadecolor} bcf & 3,144 & 23 & 3,121  &  14 & 160 & 2,984 & 0 & - & - & - & 613 & 2,531 & 445 & \textbf{3,122} & \textbf{22} & \textbf{127} \\
                      cff & 758  & 22 & 736 & 9 & 208 & 550 & 0 & - & - & - & 337 & 421 & 113 & \textbf{724} & \textbf{34} & \textbf{56} \\
\rowcolor{shadecolor} spl & 1,296 & 24 & 1,272 & 18 & 173 & 1,123 & 2 & - & - & - & 515 & 781 & 181 & \textbf{1,245} & \textbf{51} & \textbf{77} \\
                      sub & 938  & 22 & 916 & 14 & 264 & 674 & 7 & - & - & - & 393 & 545 & 236 & \textbf{885} & \textbf{53} & \textbf{127} \\ \midrule[.9pt]
\rowcolor{shadecolor} Avg.& 1,337 & 20 & 1,317 & 14 & 276 & 1,061 & 7 & 424 & 716 & 1,152 & 478 & 858 & 795 & \textbf{1,182} & \textbf{155} & \textbf{261} \\ \bottomrule[1.1pt]
\multicolumn{17}{l}{\multirow{2}{*}{\begin{tabular}[c]{@{}l@{}}\scriptsize $^*$DeepVSA's VSA implementation takes crash dumps as input and does not work on our dataset (regular binary code without crashes). Therefore, we run DeepVSA's released model on our dataset\\ and use its predicted memory region to flag dependent instructions (\S\ref{sec:impl}).
\end{tabular}}} \\
\end{tabular}
\end{table*}

% In this section, we first evaluate \sys's performance on analyzing memory dependencies (\S\ref{subsec:rq1}).
% We then perform ablation studies to show how much each design choice in \sys contribute to its performance (\S\ref{subsec:rq2}).
% Finally, we study how \sys performs on different downstream reverse engineering tasks (\S\ref{subsec:rq3}).

We focus on three main research questions in the evaluation.

\begin{itemize}[leftmargin=*]
    \item \textbf{RQ1:} How well does \sys perform in analyzing memory dependencies? (\S\ref{subsec:rq1})
    % \item \textbf{RQ2:} How fast is \sys compared to other tools?
    \item \textbf{RQ2:} How much does each design choice in \sys contribute to its performance? (\S\ref{subsec:rq2})
    %\item \textbf{RQ3:} What knowledge has \sys learned after pretrained?
    \item \textbf{RQ3:} How well does \sys perform in downstream reverse engineering tasks? (\S\ref{subsec:rq3}) % knowledge has \sys learned after pretrained?
\end{itemize}

\subsection{\sys Performance}
% \subsection{RQ1: \sys Performance in Memory Dependency Analysis}
% \RQrepeat{1}{How does \sys perform in analyzing memory dependencies?}
\label{subsec:rq1}

% We first evaluate how accurate is \sys in detecting memory dependencies and compare it to the state-of-the-art approaches.
% We then investigate how \sys generalizes to unseen binaries across various software projects, compiler optimizations, and obfuscations.

Table~\ref{tab:result} presents the results of \sys and other baselines on the test set categorized by their optimization and obfuscation flags.
\sys's results are obtained from finetuning a single model on all datasets, excluding the testing set.
On average, \sys detects 1.5$\times$ more dependencies than the second-best (DeepVSA), while having 4.5$\times$ fewer misses than the second-best tool (SVF).
We observe that Ghidra has fewer false positives than \sys, but at the cost of missing substantial dependencies, detecting 3.3$\times$ fewer dependencies than \sys.
Besides, we note that DeepVSA produces 6.4$\times$ more false positives on higher optimizations when compared to O0.
This is likely because memory access patterns (\eg using \code{rbp} and \code{rax} to access stack and heap, respectively, are largely broken by compilers.

\para{Zero-Shot Generalizability to Unseen Projects.}
In the above experiment, our training and testing set are randomly sampled with non-overlapping pairs, but they could come from the same software project.
Therefore, we investigate how \sys performs when its testing set comes from entirely different software projects.
We collect 2 software projects featuring a web server, \ie Nginx-1.21.1, and a web browser, \ie Lynx-2.8.9, to which none of the projects in our dataset has similar functionality.
We compile each software project with 4 optimizations (O0-O3), and test \sys on these unseen software projects.
As a baseline, we finetune a model where its training set includes the project, but with non-overlapping instruction pairs (``Regular'' in Table~\ref{tab:generalize}).
% We sample 5,000 positive and negative instruction pairs for each category of project and report the precision and recall of \sys.

The first two rows in Table~\ref{tab:generalize} demonstrate that \sys remains relatively robust when the testing set is collected from unseen projects, \ie on average, the number of detected dependencies only drops 1.4\% and false positives increased by 7\%. 
Interestingly, training without samples from Lynx even increases the detected dependencies, but at the expense of much higher false positives.

\begin{table}[!t]

\footnotesize
\setlength{\tabcolsep}{5pt}
\centering
\renewcommand{\arraystretch}{1.1}
\setlength\aboverulesep{0.9pt}
\setlength\belowrulesep{0.9pt}

\caption{\sys performance on binaries when they are divided with just non-overlapping train-test pairs vs. with non-overlapping programs, optimizations, and obfuscations. $\Delta$ denotes the change from regular to unseen, scaled by the number of reference dependencies.
% This tests \sys's generalizability across software projects with disparate functionality, optimizations, and obfuscations.
}
\label{tab:generalize}

\begin{tabular}{rrlllllll}
\toprule[1.1pt]
\multicolumn{2}{c}{\multirow{2}{*}{Cross-}} & \multirow{2}{*}{\# Dep} & \multicolumn{2}{c}{Regular} & \multicolumn{2}{c}{Unseen} & \multicolumn{2}{c}{$\Delta$ (+/-)} \\
 &  &  & Detect & FP & Detect & FP & Detect & FP \\ \midrule[.9pt]
\multirow{2}{*}{Proj.} & \cellcolor{shadecolor}Nginx & \cellcolor{shadecolor}226 & \cellcolor{shadecolor}164 & \cellcolor{shadecolor}56 & \cellcolor{shadecolor}155 & \cellcolor{shadecolor}68 & \cellcolor{shadecolor}-4\% & \cellcolor{shadecolor}+5.3\% \\
 & Lynx & 322 & 240 & 92 & 244 & 120 & +1.2\% & +8.7\% \\ \midrule[.9pt]
\multirow{4}{*}{Opt.} & \cellcolor{shadecolor}O0 & \cellcolor{shadecolor}1,013 & \cellcolor{shadecolor}959 & \cellcolor{shadecolor}90 & \cellcolor{shadecolor}958 & \cellcolor{shadecolor}103 & \cellcolor{shadecolor}-0.1\% & \cellcolor{shadecolor}+1.3\% \\
 & O1 & 1,310 & 1,026 & 112 & 988 & 298 & -2.9\% & +14.2\% \\
 & \cellcolor{shadecolor}O2 & \cellcolor{shadecolor}1,103 & \cellcolor{shadecolor}919 & \cellcolor{shadecolor}195 & \cellcolor{shadecolor}908 & \cellcolor{shadecolor}208 & \cellcolor{shadecolor}-1\% & \cellcolor{shadecolor}+1.2\% \\
 & O3 & 1,132 & 922 & 207 & 933 & 235 & +1\% & +2.5\% \\ \midrule[.9pt]
\multirow{4}{*}{Obf.} & \cellcolor{shadecolor}bcf & \cellcolor{shadecolor}3,144 & \cellcolor{shadecolor}3,131 & \cellcolor{shadecolor}104 & \cellcolor{shadecolor}2,946 & \cellcolor{shadecolor}925 & \cellcolor{shadecolor}-5.9\% & \cellcolor{shadecolor}+26.1\% \\
 & cff & 758 & 746 & 38 & 748 & 35 & +0.3\% & -0.4\% \\
 & \cellcolor{shadecolor}spl & \cellcolor{shadecolor}2,217 & \cellcolor{shadecolor}2,171 & \cellcolor{shadecolor}121 & \cellcolor{shadecolor}2,076 & \cellcolor{shadecolor}101 & \cellcolor{shadecolor}-4.3\% & \cellcolor{shadecolor}-0.9\% \\
 & sub & 938 & 907 & 76 & 914 & 89 & +0.8\% & +1.4\%\\ \midrule[.9pt]
\rowcolor{shadecolor}\multicolumn{2}{c}{Avg.} & 1,216 & 1,119 & 109 & 1,087 & 218 & -2.6\% & +9\% \\ \bottomrule[1.1pt]

\end{tabular}
\end{table}

\para{Zero-Shot Generalizability to Unseen Optimizations/Obfuscations.}
% We study whether \sys generalizes to unseen optimizations and obfuscations.
Aggressive compiler transformations can bring many challenges to inferring memory dependencies, \eg substituting instructions introduces more pointer arithmetic operations, which requires reasoning over the bloated instructions to detect the potential value flows.
To study whether \sys generalizes to unseen optimizations and obfuscations, we exclude binaries optimized or obfuscated by each strategy (\S\ref{sec:impl}) in training and test \sys on the excluded binaries.
% We prepare 10,000 random instruction pairs from our dataset compiled by all possible optimizations and obfuscations, excluding the obfuscation to be tested.
% We then test on 5,000 randomly sampled instruction pairs from binaries compiled by the excluded obfuscation.

Table~\ref{tab:generalize} presents results when testing \sys on each unseen optimizations and obfuscations. 
We also include the baseline results when its training set includes those optimized or obfuscated binaries (but with non-overlapping pairs). 
We observe that \sys generalizes to unseen optimizations and obfuscations, with only 2.6\% drop in detection rate and 9\% increase in false positives.

\para{Runtime Performance.}
One of the most significant benefits of \sys over traditional approaches comes from its speed, as its analysis is amenable to parallelization with GPUs.
Table~\ref{tab:speed} compares the speed of \sys to Angr and Ghidra.
We run each tool on each project compiled with O0 from our finetuning dataset (\S\ref{sec:impl}).
We observe that Angr often takes long time and cannot finish running (as also confirmed by~\cite{zhang2019bda}). 
Thus, we time it out after 5 minutes.
Consequently, Angr's actual runtime is under-estimated.
We do not compare to (i) DeepVSA because it still relies on VSA, so it is at least as slow as any VSA implementation, and (ii) SVF because it works only on LLVM IR, not directly on binaries. 
Therefore, SVF has extremely high overheard from mapping LLVM IR results to binary.
Table~\ref{tab:speed} shows that \sys is 3.5$\times$ faster than the second-best tool (Ghidra) and orders of magnitude faster (125.2$\times$) than Angr.

\begin{table}[!t]

\footnotesize
\setlength{\tabcolsep}{7pt}
\centering
\renewcommand{\arraystretch}{1.1}
\setlength\aboverulesep{0.4pt}
\setlength\belowrulesep{0.4pt}
% Dongdong: update 8 projects name.
\caption{Runtime of \sys, Angr, and Ghidra. The last column shows the speedup achieved by \sys over the second-best tool.}
\label{tab:speed}

\begin{tabular}{rrllllll}
\toprule[1.1pt]
 & \multirow{2}{*}{Size (MB)} & \multicolumn{3}{c}{Inference Time (s)} & \multirow{2}{*}{Speedup} \\ 
 &  & Angr & Ghidra & \sys \\ \midrule[.9pt]
\rowcolor{shadecolor} bash & 2.8 & 7685.9 & 60.4 & \textbf{24.4} & \textbf{2.5$\times$} \\
                      bc & 0.5 & 298.8 & 5.3 & \textbf{1.4} & \textbf{3.8$\times$} \\
\rowcolor{shadecolor} binutils & 74 & 70157.1 & 3077.2 & \textbf{695.6} & \textbf{4.4$\times$} \\
                      bison &  1.6 & 1730.1 & 30.2 & \textbf{10.3} & \textbf{2.9$\times$} \\ 
\rowcolor{shadecolor} cflow & 0.56 & 695.9 & 6.5 & \textbf{3} & \textbf{2.2$\times$} \\ 
                      coreutils & 16 & 40,188.2 & 392.2 & \textbf{105.9} & \textbf{3.7$\times$} \\ 
\rowcolor{shadecolor} curl & 0.77 & 91.5 & 14.5 & \textbf{3.1} & \textbf{4.7$\times$} \\ 
                      findutils & 2.3 & 882.7 & 80.1 & \textbf{23.2} & \textbf{3.5$\times$} \\  
\rowcolor{shadecolor} gawk & 3.8 & 2305.3 & 55.0 & \textbf{14.1} & \textbf{3.9$\times$} \\ \midrule[.9pt]
                      Avg. &  12.8 & 13781.7 & 413.5 & \textbf{110.1} & \textbf{3.5$\times$} \\ \bottomrule[1.1pt]
\end{tabular}
\end{table}

\subsection{Ablation Study}
% \RQrepeat{2}{How much does each design choice in \sys contribute to its performance?}
%\subsection{RQ2.~How much does each design choice in NEUDEP contribute to its performance?}
\label{subsec:rq2}

We study how much each design in \sys (\S\ref{sec:method}) contributes to its performance.
We follow the setup in \S\ref{subsec:rq1}.
% With each design, we pretrain and finetune like how we train the model in \S\ref{subsec:rq1}, and test on the test set in Table~\ref{tab:result}.
% Therefore, the total number of reference dependencies are 10,694.
Table~\ref{tab:ablation} summarizes the results where we bold \sys's default choice.
% We start from comparing \sys's performance when it is pretrained with reasoning over execution to that without pretraining.
% We then ablate design specifics, such as the strategy of aggregating byte sequence, fusing inputs, and compare to the default choice with alternatives when necessary.
% We consider 5 key designs in \sys that potentially influence \sys performance, described in the following.
\begin{table}[!t]

\footnotesize
\setlength{\tabcolsep}{3.5pt}
\centering
\renewcommand{\arraystretch}{1.1}
\setlength\aboverulesep{0.4pt}
\setlength\belowrulesep{0.4pt}
% Dongdong: update 8 projects name.
\caption{Ablation on \sys designs. We treat the first row of each design as the baseline and compute the improvement of other alternatives. }
\label{tab:ablation}

\begin{tabular}{rrllllll}
\toprule[1.1pt]
\multicolumn{2}{c}{\multirow{2}{*}{Ablation Setup}} & \multirow{2}{*}{Detect} & \multirow{2}{*}{Miss} & \multirow{2}{*}{FP} & \multicolumn{3}{c}{Improve (+/-)} \\ 
 &  &  &  &  & Detect & Miss & FP \\ \midrule[.9pt]
\multirow{2}{*}{Pretrain} & \cellcolor{shadecolor}w/o & \cellcolor{shadecolor}8,780 & \cellcolor{shadecolor}1,914 & \cellcolor{shadecolor}1,477 & \cellcolor{shadecolor}0.0\% & \cellcolor{shadecolor}0.0\% & \cellcolor{shadecolor}0.0\% \\
 & \textbf{w/} & \textbf{9,882} & \textbf{812} & \textbf{1,013} & \textbf{+12.6\%} & \textbf{-57.6\%} & \textbf{-31.4\%} \\ \midrule[.9pt]
% \multicolumn{5}{c}{Pretraining Objectives} \\ 
% \rowcolor{shadecolor} Interpret &  &  &  & 0.0\% \\
% Synthesize &  &  &  &  \\
% \rowcolor{shadecolor} Interpret+Synthesize &  &  &  &  \\ \midrule[.9pt]
\multirow{2}{*}{\begin{tabular}[r]{@{}r@{}}Value\\ Embed\end{tabular}} & \cellcolor{shadecolor}Concat & \cellcolor{shadecolor}9,666 & \cellcolor{shadecolor}1,208 & \cellcolor{shadecolor}1,027 & \cellcolor{shadecolor}0.0\% & \cellcolor{shadecolor}0.0\% & \cellcolor{shadecolor}0.0\% \\
 & \textbf{Conv.} & \textbf{9,882} & \textbf{812} & \textbf{1,013} & \textbf{+2.2\%} & \textbf{-32.8\%} & \textbf{-1.4\%} \\ \midrule[.9pt]
\multirow{4}{*}{\begin{tabular}[r]{@{}r@{}}Fusing\\ Strategy\end{tabular}} & \cellcolor{shadecolor}Sum & \cellcolor{shadecolor}9,752 & \cellcolor{shadecolor}942 & \cellcolor{shadecolor}1,419 & \cellcolor{shadecolor}0.0\% & \cellcolor{shadecolor}0.0\% & \cellcolor{shadecolor}0.0\% \\
 & \textbf{1st Layer} & \textbf{9,882} & \textbf{812} & \textbf{1,013} & \textbf{+1.3\%} & \textbf{-13.8\%} & \textbf{-28.6\%} \\
 & \cellcolor{shadecolor}3rd Layer & \cellcolor{shadecolor}9,870 & \cellcolor{shadecolor}824 & \cellcolor{shadecolor}1,209 & \cellcolor{shadecolor}+1.2\% & \cellcolor{shadecolor}-12.5\% & \cellcolor{shadecolor}-14.7\% \\
 & 5th Layer & 9,700 & 994 & 1,186 & +0.5\% & -5.5\% & -16.4\% \\ \midrule[.9pt]
\multirow{2}{*}{\begin{tabular}[r]{@{}r@{}}Compos.\\ Learning\end{tabular}} & \cellcolor{shadecolor}w/o Compos. & \cellcolor{shadecolor}9,806 & \cellcolor{shadecolor}888 & \cellcolor{shadecolor}1,389 & \cellcolor{shadecolor}0.0\% & \cellcolor{shadecolor}0.0\% & \cellcolor{shadecolor}0.0\% \\ 
 & \textbf{w/ Compos.} & \textbf{9,882} & \textbf{812} & \textbf{1,013} & \textbf{+0.8\%} & \textbf{-8.6\%} & \textbf{-27.1\%} \\ \midrule[.9pt]
\multirow{2}{*}{\begin{tabular}[r]{@{}r@{}}Code\\ Addr.\end{tabular}} & \cellcolor{shadecolor}w/o Addr. & \cellcolor{shadecolor}9,667 & \cellcolor{shadecolor}1,027 & \cellcolor{shadecolor}1,407 & \cellcolor{shadecolor}0.0\% & \cellcolor{shadecolor}0.0\% & \cellcolor{shadecolor}0.0\% \\ 
 & \textbf{w/ Addr.} & \textbf{9,882} & \textbf{812} & \textbf{1,013} & \textbf{+2.2\%} & \textbf{-20.9\%} & \textbf{-28\%} \\ \bottomrule[1.1pt]
\end{tabular}
\end{table}

\para{Pretraining.}
% The key idea of designing our pretraining strategy is that it compels the model to infer instructions compositional behavior and perform flow and path sensitive memory dependence analysis.
We first ablate the effectiveness of pretraining in assisting memory dependence analysis.
Table~\ref{tab:ablation} shows that pretraining \sys significantly improves its performance by 12.6\% in the number of detected dependencies.
The number of misses and false positives drop by 57.6\% and 31.4\%, respectively.

\para{Byte Aggregation.}
We study the effectiveness of the design to encode numeric values using convolutions with highway network (\S\ref{subsec:inputs}) by comparing it to the baseline that concatenates the input bytes.
Table~\ref{tab:ablation} shows that our encoding mechanism outperforms the baseline by 2.2\% and significantly reduces the miss detection rate by 32.8\%.
% Saying that, encoding numeracy into the model is still a challenging task~\cite{Wallace2019Numbers}. We envision an exciting future work to study various encoding mechanims of byte sequences and architectures that can efficiently learn interactions between instructions and encoded byte values.
% To study what design is the most effective to learn numeric representations of byte sequences, we ablate 4 strategies: 
% \begin{itemize}[leftmargin=*]
%     \item Vector sum of the embeddings of each byte.
%     \item Concatenate each byte embeddings and reduce the resulting increased dimensions back to original dimension.
%     \item Employing multiple Convolutional filters with different receptive fields (for different sub-sequences) and aggregate them with a highway network.
% \end{itemize}

\para{Input Fusion.}
We explore the effectiveness of input fusion by comparing it to the baseline that takes the vector sum of the input embeddings~\cite{pei2020trex, pei2021stateformer}.
We also study fusing after which layer is the most effective. 
We note that simply summing the embeddings of code and trace values at input by assuming they are homogeneous performs the worst.
This confirms our intuition that code and trace are heterogeneous data that benefit from different encoding mechanisms. 
In addition, we note that combining code and trace at earlier layers performs the best.
This is likely because trace values can participate early in the model's computation of interactions between instructions and trace values, \ie fusing in the later layers implies it has fewer remaining layers to learn how code and trace interacts.

\para{Composition Learning.}
We study whether composition learning (\S\ref{subsec:overview_design}) would bring any positive effect on the model's finetuning performance for detecting memory dependencies.
We compare it to the fixed masking percentage strategy where the samples are shuffled randomly, and the masking rate $r$ is fixed to 0.5 on both the code and trace tokens.
Table~\ref{tab:ablation} shows that composition learning moderately improves the model by 0.8\% in detected dependencies but substantially reduces the number of false positives by 27.1\%.
This observation confirms our intuition that arranging the training samples based on their difficulty helps the model learn more efficiently.

\para{Modeling Address Layout.}
We study whether annotating the binary code with its loaded addresses would bring a useful inductive bias to the model by comparing to the baselines that do not model them~\cite{pei2020trex, pei2021stateformer}.
Table~\ref{tab:ablation} shows that annotating the code with addresses significantly reduces the model's missed detection and false positives, \ie by 20.9\% and 28\%, respectively.
This shows that the code address helps the model reduce the spurious dependencies.

% Table~\ref{tab:ablation} summarizes the results of ablation studies. 
% First, it validates the assumption that pretraining with both assembly code and execution traces is effective: \sys pretrained with such a strategy improves on that without pretraining by XX\%.
% It also outperforms the pretrained model on only code by XX\%.
% The empirical evidence justifies our claim -- unlike natural language, code semantics is not manifested by its neighboring context, but grounded by how it executes.

%\subsection{RQ3: Probing Learned Knowledge}
\subsection{Performance on Reverse Engineering Tasks}
% \RQrepeat{3}{How does \sys perform in downstream reverse engineering tasks?}
\label{subsec:rq3}

% \para{Direct calls with target removed.}
% \sys is primarily designed for robust performance on downstream memory dependence tasks, but it is important to keep in mind that these tasks do not exist in a vacuum. Many other important binary analysis tasks may benefit heavily from knowledge of memory dependence. In this section, we demonstrate that \sys's pretrained knowledge can be easily transferred to other downstream applications, producing state-of-the-art results. These diverse downstream tasks also serve as natural probing tasks, and they can give us some insight into the knowledge that \sys has captured. 
We probe pretrained \sys using three reverse engineering tasks that either assist or benefit from memory dependence analysis.
% The performance on these diverse downstream tasks can help the knowledge that \sys has captured. 
% As defined in \S\ref{subsec:probing_tasks_overview}, these tasks are 

\para{Memory-Access Regions.}
We follow PalmTree by running \sys on the \emph{DeepVSA's dataset} and compare \sys to the reported F1 scores of PalmTree, DeepVSA, and other baselines (\S\ref{sec:impl}).
We note that DeepVSA's datasets are all 32-bit x86 binaries, but \sys is pretrained on x86-64 binaries. 
However, we find that just our vocabulary constructed from x86-64 binaries covers 89.9\% of DeepVSA's dataset vocabulary, likely because both belong to the x86 family.
% As we have to keep the same vocabulary when using the pretrained model, 
Therefore, we simply apply our vocabulary on DeepVSA's dataset and replace unseen tokens with ``unknown'' in the vocabulary.
% As DeepVSA has compared to other alternative neural networks, \ie Hidden-Markov Model (HMM), Conditional Random Field (CRF), Bidirectional Recurrent Neural Net (Bi-RNN), Bidirectional Gated Recurrent Unit (Bi-GRU), Bidirectional Long-Short Term Memory (Bi-LSTM), we also include their results as part of our baselines.
% Let $R$ denotes recall, which measures out of the chosen groundtruth label, how many of them are correctly detected.
% Let $P$ denotes the precision, which measures out of the predicted labels, how many of them are correctly predicted.
% F1 score is computed by $2\cdot P\cdot R/(P+R)$.

Table~\ref{tab:deepvsa} shows that \sys remains robust across different memory regions.
On average, \sys outperforms PalmTree by 0.069.
On more challenging labels such as heap, \sys outperforms PalmTree and DeepVSA by 0.19 and 0.32, respectively.
This is likely because accessing these memory regions involves more diverse patterns, \eg via the stack pointer register (Figure~\ref{fig:task_definition}).

\begin{table}[!t]

\footnotesize
\setlength{\tabcolsep}{8pt}
\centering
\renewcommand{\arraystretch}{1.1}
\setlength\aboverulesep{0.4pt}
\setlength\belowrulesep{0.4pt}
% Dongdong: update 8 projects name.
\caption{Comparison of F1 scores on memory region prediction between \sys and PalmTree its other studied baselines. }
\label{tab:deepvsa}

\begin{tabular}{rllllr}
\toprule[1.1pt]
    &  Global & Heap & Stack & Other & Avg. \\ \midrule[.9pt]
\rowcolor{shadecolor} Instruction2Vec & 0.654 & 0.566 & 0.914 & 0.947 & 0.77 \\
                      Asm2Vec & 0.517 & 0.359 & 0.911 & 0.948 & 0.684 \\
\rowcolor{shadecolor} DeepVSA & 0.835 & 0.584 & 0.944 & 0.959 & 0.831 \\
                      PalmTree & 0.855 & 0.714 & 0.95 & 0.971 & 0.873 \\ \midrule[.9pt]
\rowcolor{shadecolor} \sys & \textbf{0.91} & \textbf{0.904} & \textbf{0.977} & \textbf{0.976} & \textbf{0.942} \\ \bottomrule[1.1pt]
\end{tabular}
\end{table}

\para{Function Signature.} 
We compare \sys to EKLAVYA on recovering function signatures. 
Table~\ref{tab:signature} shows that \sys outperforms EKLAVYA on all signature inference tasks, achieving 12.6\% higher accuracy on average. 
Most notably, \sys's performance remains robust across different tasks and optimization levels, while EKLAVYA's accuracy shows clear drops. 
For instance, when comparing the prediction accuracy of 3rd argument ($A_3$) and 1st argument ($A_1$), EKLAVYA decreases by 16.61\% while \sys's drops by only 1.19\%. 
Likewise, within the arity task, EKLAVYA's accuracy decreases 14.02\% from O0 to O3, while \sys decreases 2.91\%. 
% This is possibly because EKLAVYA is not pretrained via execution traces. 
% Thus, as the complexity of the code increases with higher optimization levels \todo{how can we fit argument type position in this idea as well?}, EKLAVYA becomes less able to parse the execution semantics.

\begin{table}[!t]

\footnotesize
\setlength{\tabcolsep}{4pt}
\centering
\renewcommand{\arraystretch}{1.2}
\setlength\aboverulesep{0.4pt}
\setlength\belowrulesep{0.4pt}
\caption{We compare \sys to EKLAVYA's accuracy on five function signature tasks across 4 optimizations for caller and callee. 
% \todo{need to mention how EKLAVYA values are obtained? ie self reported vs our training}
% This tests \sys's ability to infer information about memory dependence from different input forms and its robustness across several optimization levels.
}
\label{tab:signature}

\begin{tabular}{rccccccccc}

\toprule[1.1pt] \multicolumn{1}{l}{} & & \multicolumn{4}{c}{Caller} & \multicolumn{4}{c}{Callee}  \\
 \multicolumn{1}{l}{} & & O0 & O1 & O2 & O3 & O0 & O1 & O2 & O3\\ \midrule[.9pt]
\parbox[t]{2mm}{\multirow{2}{*}{\rotatebox[origin=c]{90}{Ret.}}} & \multicolumn{1}{c}{EKLA.} & 66.62 & 70.59 & 73.63 & 76.19 & 91.59 & 88.87 & 91.92 & 95.32 \\
&  \multicolumn{1}{c}{\cellcolor{shadecolor}\sys} & \cellcolor{shadecolor}\textbf{94.65} & \cellcolor{shadecolor}\textbf{93.33} & \cellcolor{shadecolor}\textbf{95.75} & \cellcolor{shadecolor}\textbf{96.41} & \cellcolor{shadecolor}\textbf{95.37} & \cellcolor{shadecolor}\textbf{93.42} & \cellcolor{shadecolor}\textbf{96.06} & \cellcolor{shadecolor}\textbf{98.20} \\
\midrule[.9pt]
\parbox[t]{2mm}{\multirow{2}{*}{\rotatebox[origin=c]{90}{$A_1$}}} & \multicolumn{1}{c}{EKLA.} & 91.56 & 90.38 & 91.21 & 91.55 & 95.62 & 92.40 & 93.05 & 92.56 \\
&  \multicolumn{1}{c}{\cellcolor{shadecolor}\sys} & \cellcolor{shadecolor}\textbf{97.03} & \cellcolor{shadecolor}\textbf{97.09} & \cellcolor{shadecolor}\textbf{98.47} & \cellcolor{shadecolor}\textbf{99.17} & \cellcolor{shadecolor}\textbf{97.24} & \cellcolor{shadecolor}\textbf{95.10} & \cellcolor{shadecolor}\textbf{97.01} & \cellcolor{shadecolor}\textbf{97.84} \\
\midrule[.9pt]
\parbox[t]{2mm}{\multirow{2}{*}{\rotatebox[origin=c]{90}{$A_2$}}} & \multicolumn{1}{c}{EKLA.} & 81.82 & 78.70 & 81.81 & 82.03 & 87.25 & 82.67 & 82.40 & 85.07 \\
&  \multicolumn{1}{c}{\cellcolor{shadecolor}\sys} & \cellcolor{shadecolor}\textbf{96.08} & \cellcolor{shadecolor}\textbf{94.86} & \cellcolor{shadecolor}\textbf{97.68} & \cellcolor{shadecolor}\textbf{97.51} & \cellcolor{shadecolor}\textbf{93.30} & \cellcolor{shadecolor}\textbf{91.34} & \cellcolor{shadecolor}\textbf{94.66} & \cellcolor{shadecolor}\textbf{92.45}\\
\midrule[.9pt]
\parbox[t]{2mm}{\multirow{2}{*}{\rotatebox[origin=c]{90}{$A_3$}}} & \multicolumn{1}{c}{EKLA.} & 80.28 & 79.85 & 81.35 & 76.63 & 77.42 & 69.18 & 70.93 & 69.80 \\
&  \multicolumn{1}{c}{\cellcolor{shadecolor}\sys} & \cellcolor{shadecolor}\textbf{96.88}& \cellcolor{shadecolor}\textbf{96.89} & \cellcolor{shadecolor}\textbf{97.09} & \cellcolor{shadecolor}\textbf{97.24} & \cellcolor{shadecolor}\textbf{96.55} & \cellcolor{shadecolor}\textbf{94.42} & \cellcolor{shadecolor}\textbf{94.66} & \cellcolor{shadecolor}\textbf{95.68} \\
\midrule[.9pt]
\parbox[t]{2mm}{\multirow{2}{*}{\rotatebox[origin=c]{90}{Arity}}} & \multicolumn{1}{c}{EKLA.} & 92.03 & 86.02 & 83.80 & 82.79 & 97.48 & 76.24 & 77.49 & 78.69 \\
&  \multicolumn{1}{c}{\cellcolor{shadecolor}\sys} & \cellcolor{shadecolor}\textbf{98.84}& \cellcolor{shadecolor}\textbf{95.44} & \cellcolor{shadecolor}\textbf{96.35} & \cellcolor{shadecolor}\textbf{95.86} & \cellcolor{shadecolor}\textbf{99.23} & \cellcolor{shadecolor}\textbf{92.57} & \cellcolor{shadecolor}\textbf{95.04} & \cellcolor{shadecolor}\textbf{96.40} \\

\bottomrule[1.1pt]
\end{tabular}
\end{table}

% \para{Negative control.} 
% As argued above, using a naive aggregation strategy for the caller tasks creates an impossible task for the model. The input is not meaningful, and the model has no way of even knowing what it is supposed to predict. Therefore, we can use this impossible setup as a negative control. If \sys is simply memorizing answers, then it should be able to predict with high accuracy regardless of whether the input contains meaningful information or not. On the contrary, if it needs to perform some meaningful internal inference to produce its predictions, then we expect \sys to fail on this task. 

\para{Indirect Calls.}
% To further demonstrate the robustness of \sys's end-to-end indirect call prediction, we show evidence that \sys's ability to predict function signatures can be directly leveraged to perform caller-callee matching. 
Finally, we compare how well \sys, EKLAVYA, and TypeArmor detect indirect calls (Definition~\ref{def:indirect_call_prediction}). 
We consider 8 matching algorithms (\S\ref{subsec:probing_methodology}) grouped row-wise by arity matching criteria detailed in Table~\ref{tab:indirect_call}. 
On all algorithms, \sys outperforms EKLAVYA and TypeArmor, achieving 0.032 and 0.07 higher F1 scores, respectively. 
With loose arity and return type matching -- the criterion adopted in TypeArmor -- \sys outperforms TypeArmor by 0.052 in F1 score. 
We also note that \sys's performance increases as the matching algorithm incorporates more conditions, while the performance of other systems remains roughly the same. 
% We conjecture that this may be due to \sys's high signature inference accuracy: incorporating more matching conditions (and thus more components of the signature) allows for tighter matching, provided that all components are inferred correctly. When components are inferred inaccurately, incorporating more components introduces more chances for errors.

\begin{table}[!t]

\footnotesize
\setlength{\tabcolsep}{5pt}
\centering
\renewcommand{\arraystretch}{1.5}
\setlength\aboverulesep{0.4pt}
\setlength\belowrulesep{0.4pt}
\caption{\sys's performance (F1 score) on matching indirect calls using several heuristic algorithms. TypeArmor cannot infer argument types, so the corresponding cells are dashed out.}
\label{tab:indirect_call}

\begin{tabular}{ccrrrrrrrr}
\toprule[1.1pt]  & & Arity & Arity+Ret & \hphantom{5}Arity+Arg & \hphantom{5}Arity+Arg+Ret\\
\midrule[.9pt] 
\parbox[t]{2mm}{\multirow{3}{*}{\rotatebox[origin=c]{90}{\makecell{Loose}}}} & TypeArmor & 0.75 & 0.752 & - & -\\
& \cellcolor{shadecolor}EKLAVYA & \cellcolor{shadecolor}0.777 & \cellcolor{shadecolor}0.778 & \cellcolor{shadecolor}0.8 & \cellcolor{shadecolor}0.801  \\
& \sys & \textbf{0.783} & \textbf{0.804}& \textbf{0.83} & \textbf{0.843}\\

\midrule[.9pt]
\parbox[t]{2mm}{\multirow{3}{*}{\rotatebox[origin=c]{90}{\makecell{Strict}}}} & \cellcolor{shadecolor}TypeArmor & \cellcolor{shadecolor}0.777 & \cellcolor{shadecolor}0.778 & \cellcolor{shadecolor}- & \cellcolor{shadecolor}- \\
& EKLAVYA & 0.818 & 0.817 & 0.811 & 0.811\\
& \cellcolor{shadecolor}\sys & \cellcolor{shadecolor}\textbf{0.844} & \cellcolor{shadecolor}\textbf{0.853} & \cellcolor{shadecolor}\textbf{0.851} & \cellcolor{shadecolor}\textbf{0.857}\\
\bottomrule[1.1pt]
\end{tabular}

\end{table}

\section{Threats to Validity}
\label{sec:threat}

\para{Architecture Bias.} We only consider x86-64 binaries.
While we have shown \sys generalizes to several x86-32 binaries (\S\ref{subsec:rq3}), it cannot directly be applied to binaries with significantly different syntax, \eg firmware usually run on ARM or MIPS architectures.
However, given the fact that our trace engine supports other architectures well~\cite{qiling}, we can potentially pretrain the model for other architectures.
We also plan to extend our models to different programming languages that come with efficient tracing support~\cite{pina2016tedsuto,grimmer2014trufflec,mondal2021soundy}.

\para{Performance Bias.}
We only compare \sys's runtime performance on GPUs with other baselines (\S\ref{subsec:rq1}), as \sys's neural module runs on GPU by default.
However, we believe that significantly benefitting from GPU is indeed a key advantage of ML-based techniques over traditional binary analysis that cannot easily exploit GPU parallelism and thus struggle to scale to large binaries.

\para{Ground Truth Bias.}
Obtaining complete ground truth for memory dependencies in real-world programs is intractable.
Therefore, following BDA's approach~\cite{zhang2019bda}, we resort to dynamic analysis and use the accessed memory locations observed during execution to collect the reference dependencies.
While we cannot guarantee the ground truth to be complete, this approach can still quantify how many dependencies are missed by the evaluated tools.
Table~\ref{tab:result} shows that \sys outperforms all baselines with the fewest misses.

\para{Inter-Procedural Analysis.}
In our experiment, we only capture full execution behavior starting from a callee. 
Therefore, \sys primarily expects instruction pairs to come from the same function. 
However, as we trace the full execution behavior of callee, our model potentially learns to reason about the value flows across procedures.
We plan to explore \sys's capability in generic inter-procedural memory dependence by modeling the complete calling context in our future study.

\begin{comment}
\para{Case Studies on Crashes or Vulnerabilities.} 
While we have shown \sys's strong generalizability to large binaries whose reliability and security are critical, \eg web servers and browsers (\S\ref{subsec:rq1}), we do not perform case studies on how \sys helps to identify the root cause of critical crashes.
However, as we have shown that \sys outperforms DeepVSA in predicting memory regions (\S\ref{subsec:rq3}), we are confident that replacing the neural network in DeepVSA with \sys can help instantiate a more accurate value set for VSA to analyze the crashes (\eg with less false positives).
We leave this study as our future work.
\end{comment}

\section{Related Work}
\label{sec:related}

\para{Binary Memory Dependence Analysis.}
There has been a long history of efforts to approach the problem of analyzing memory dependencies in executables~\cite{debray1998alias, cifuentes1997intraprocedural, guo2005practical,brumley2006alias, balakrishnan2004vsa, balakrishnan2010wysinwyx, reps2008improved, guo2019vsa, zhang2019bda}. 
Debray~\etal\cite{debray1998alias} and Cifuentes~\etal\cite{cifuentes1997intraprocedural} pioneered this field by using abstract interpretation to propagate the abstract domain along the registers of each instruction.
VSA~\cite{balakrishnan2004vsa} improves on their idea by supporting tracking value flows along both the registers and memory locations.
DeepVSA~\cite{guo2019vsa} further improves on VSA by learning a neural network to predict the memory-access regions of each instruction to pre-filter those not sharing the regions.
BDA~\cite{zhang2019bda} uses probabilistic analysis to uniformly sample paths and performs per-path abstract interpretation to avoid precision losses from path merging. 
While both DeepVSA and BDA sacrifice soundness, they have been shown to significantly assist in debugging crashes~\cite{cui2018rept, mu2019renn} and malware analysis.
However, they still incur high runtime overhead and produce many false positives for optimized binaries -- \sys substantially outperforms these tools (\S\ref{sec:eval}).
% There have also been efforts in applying machine learning to VSA. 
% Specifically, DeepVSA ~\cite{guo2019vsa} uses an LSTM to determine the different types of memory regions in VSA. Spindle~\cite{wang2018spindle} uses backward data flow analysis to build a memory dependence tree for memory accesses at compile time. 
% Jeon~\etal\cite{jeon2007layout} uses a regular expression approach at compile time to determine memory access patterns. 
% Lin~\etal\cite{lin2019value} combines value set analysis with conditional merging and lazy constraint solving to speed up performance on value set analysis for memory dependencies(I NEED TO READ THIS PAPER BETTER). 
%However, similar to using VSA for indirect calls, using VSA for Binary Dependence Analysis can be prohibitively expensive.

\para{Machine Learning for Program Analysis.}
Machine learning has been shown to be extremely promising in analyzing both source code and executables~\cite{sharma2021survey,devanbu2020deep,puasuareanu2012learning,allamanis2018survey} in tasks like type inference~\cite{hellendoorn2018deep,pandi2021type,pradel2020typewriter,pei2021stateformer,xu2016python,he2018debin,mir2021type4py}, code completion~\cite{izadi2022codefill,ciniselli2021empirical,bhoopchand2016learning}, program synthesis and generation~\cite{Wang2021DataDrivenSO,sun2019grammar}, program repair and fix~\cite{wang2017dynamic,svyatkovskiy2021mergebert,dinella2021deepmerge,zhu2021syntax,allamanis2014learning,huq2022review4repair}, code summarization~\cite{shia2022evaluation,bui2021self,david2020neural,wang2019learning}, general code representation~\cite{ma2021graphcode2vec,li2021palmtree,wang2020blended,bui2021infercode,hindle2016naturalness}, bug/vulnerability detection~\cite{patra2021nalin,sellik2021learning,degiovanni2022mu,wang2020combining,kim2019precise}, code clone detection and search~\cite{ishtiaq2021bert2code,gui2022cross, bui2021self,gu2021cradle,mehrotra2021modeling,pei2020trex}, code translation~\cite{roziere2021leveraging}, comment suggestion~\cite{louis2020should,huang2019learning}, and reverse engineering tasks~\cite{pei2020xda,menguy2021ai,bardin2021compiler,benoit2021binary}.
% At the binary level, ~\cite{li2021palmtree, koo2021semantic, banerjee2021variable,pei2020trex,pei2020xda,pei2021stateformer} each pretrain embeddings of binary code for downstream tasks such as disassembly, type inference, and binary similarity. 
% At the source level, LambdaNet~\cite{wei2020lambdanet} and PYInfer~\cite{Cui2021PYInferDL} infer the variable types using a neural network model.
% Chae~\etal\cite{Chae2017AutomaticallyGF} proposes to automatically learn reduced and abstract program features for various data-driven program anlaysis tasks. 
% Cito~\etal\cite{Cito2021CounterfactualEF} tries to improve the explainability of source code model by integrating counterfactual explanation to the model output. 
% Wang~\etal\cite{Wang2021DataDrivenSO} leverages decision tree learning to synthesize static analyses for detecting side-channel information leak. 
% Code2Inv~\cite{si2020code2inv} learns valid proof for program validation task in an end-to-end manner.
Recent works have observed that incorporating program behavior is beneficial to learning more effective program representations~\cite{pei2021stateformer, nye2021show, pei2020trex, wang2017dynamic, wang2020blended, patra2021nalin}.
For example, Pei~\etal\cite{pei2021stateformer, pei2020trex} demonstrate that pretraining ML models with execution traces can help the model understand the program's operational semantics, showing successes in detecting semantically similar binaries and type inference under various code transformations~\cite{zhang2021challenging}.
However, they have no support for data flow through memory and thus do not model fine-grained value flows across memory operations.
We show in \S\ref{subsec:rq2} that \sys's new designs absent in these works are critical to analyzing memory dependencies.

% Recent successes in self-supervised learning of using large pretrained neural net has transformed the paradigm
% Pretraining-finetuning has been a new paradigm, and people have shifted the attention from designing new network architecture with specified inductive bias to designing new training objectives.
% The idea is that the required knowledge emerge during the process of optimizing towards these training objectives.
 %~\cite{li2021palmtree, koo2021semantic, banerjee2021variable, pei2020xda, pei2020trex, pei2021stateformer, ahmad2020summarization, feng2020codebert}

\section{Conclusion}
\label{sec:conclusion}

We present a new ML-based approach, \sys, to predict memory dependencies.
We first pretrain \sys to understand how instructions propagate dynamic values across memory and registers, then finetune the model to detect memory dependencies statically.
We demonstrate that \sys is precise and efficient, outperforming the state-of-the-art in both detection accuracy (1.5$\times$) and speed (3.5$\times$).
Extensive probing studies demonstrate that \sys understands memory access patterns, learns function signatures, and can match indirect calls -- these tasks either assist or benefit from inferring memory dependencies. 
Notably, \sys also outperforms the state-of-the-art on these tasks.

\begin{acks}
We thank the anonymous reviewers for their constructive and valuable feedback. 
We thank the author of BDA, Zhuo Zhang, for providing valuable insight and suggestion, and running experiments of BDA.
This work is sponsored in part by NSF grants CCF-1845893, CCF-2107405, CNS-1564055, and IIS-2221943; ONR grant N00014-17-1-2788; an NSF Career Award; an Accenture Faculty Research Award; a Google Gift; an IBM Faculty Award. 
Any opinions, findings, conclusions, or recommendations expressed herein are those of the authors, and do not necessarily reflect those of the US Government, NSF, ONR, Accenture, Google, or IBM.
\end{acks}

\clearpage
\bibliographystyle{ACM-Reference-Format}
\bibliography{ref}

\end{document}